\RequirePackage{rotating}
\RequirePackage{lineno}
\PassOptionsToPackage{usenames,dvipsnames}{xcolour}

\documentclass[a4paper,twocolumn, numberedappendix, twocolappendix,revtex4,iop]{openjournal} 
\usepackage[a4paper,marginpar=10pt,headheight=20pt, right=1cm, top=1.5in, left=1in, bottom=0pt, footskip=0pt, footnotesep=10pt]{geometry}

\pdfoutput=1 %for arXiv submission
\usepackage{amsmath,amstext}
\usepackage[T1]{fontenc}
\usepackage{apjfonts}
\usepackage{ae,aecompl}
\usepackage[utf8]{inputenc}
\usepackage{natbib}
\usepackage{url}
\usepackage{mdwlist}
\usepackage{listings}
\usepackage{rotating}
\usepackage{multirow}
\usepackage{times}
\usepackage{booktabs}
\usepackage{multirow}
\usepackage{placeins}

\usepackage[colorlinks,linkcolor=blue,citecolor=blue,urlcolor=blue ]{hyperref}
\usepackage{xurl} %to allow line breaks on urls that are too long
\usepackage{xspace}
\usepackage{pifont}
\usepackage{orcidlink}
\newcommand{\cmark}{\ding{51}}%
\newcommand{\xmark}{\ding{55}}%

\usepackage{xcolor}          % Enables coloured text
\definecolor{orange}{RGB}{255,132,0}
\definecolor{pink}{RGB}{249,135,197}
 
\definecolor{gold}{RGB}{180,132,85}

\usepackage{ulem}
\usepackage{comment}

\newcommand{\bb}[1]{\left[ #1 \right]}

\defcitealias{des/kids:2023}{DK23}
\defcitealias{DESCSRD/etal:2018}{DESC-SRD}
\newcommand{\Euclid}{{\it Euclid}\xspace}
\newcommand{\Rubin}{{\it Rubin}\xspace}
\newcommand{\hmcode}{{\sc HMCode2020}\xspace}
\newcommand{\cosmosis}{{\sc CosmoSIS}\xspace}
\newcommand{\cosmopower}{{\sc CosmoPower}\xspace}
\newcommand{\halofit}{{\sc HALOFIT}\xspace}
\newcommand{\camb}{{\sc CAMB}\xspace}
\newcommand{\spk}{{\sc SP(k)}\xspace}
\newcommand{\nautilus}{{\sc Nautilus}\xspace}
\newcommand{\eg}{\textit{e.g.}}

\shorttitle{Impact and interplay of $\Lambda$CDM analysis choices for LSST cosmic shear}
\shortauthors{Robertson et al.\ (LSST~DESC)}

\begin{document}
\title{Impact and interplay of $\Lambda$CDM analysis choices for LSST cosmic shear}
\author{\vspace*{-35px}N. C. Robertson\orcidlink{0009-0006-3709-7581}\altaffilmark{1*}, C. Heymans\orcidlink{0000-0001-7023-3940}\altaffilmark{1,2}, 
J. Zuntz\orcidlink{0000-0001-9789-9646}\altaffilmark{1}, 
P. Burger \altaffilmark{3,4}, 
C. D. Leonard\orcidlink{0000-0002-7810-6134} \altaffilmark{5}
I. G. McCarthy\orcidlink{0000-0002-1286-483X} \altaffilmark{6}, 
J. G. Paine \altaffilmark{7}, 
J. Salcido\orcidlink{0000-0002-8918-5229} \altaffilmark{6}, 
N. {\v{S}}ar{\v{c}}evi{\'c}\orcidlink{0000-0001-7301-6415} \altaffilmark{8},
M. Schaller\orcidlink{0000-0002-2395-4902} \altaffilmark{9,10},
J. Schaye\orcidlink{0000-0002-0668-5560} \altaffilmark{10},
M. P. van Daalen\orcidlink{0000-0002-8801-4911} \altaffilmark{10}
the FLAMINGO team and The LSST Dark Energy Science Collaboration.\\
\textit{Author affiliations may be found before the references.}}
\email[$^\ast$ Email:]{naomi.robertson@ed.ac.uk}

%\collaboration{The LSST Dark Energy Science Collaboration, LSST DESC}

\begin{abstract}
We forecast cosmological parameter constraints for a cosmic shear analysis of the Rubin Observatory Legacy Survey of Space and Time (LSST), defining an analysis framework that can accurately recover the $\Lambda$CDM model in the presence of astrophysical and data-related systematics.
When accounting for our present uncertainty on the suppression of the non-linear matter power spectrum through baryon feedback, we find that the error on the composite parameter $S_8=\sigma_8\sqrt{\Omega_{\rm m}/0.3}$ almost doubles compared to an LSST analysis which neglects this astrophysical phenomenon.
After the first year of observations, LSST will extend beyond the magnitude limit of existing representative spectroscopic calibration samples, requiring photometric redshifts to be calibrated using an alternative strategy.  Adopting literature measurements of the reduced redshift calibration precision found from galaxy cross-correlation techniques, combined with current levels of baryon feedback uncertainty, we forecast final year LSST cosmic shear constraints that barely improve upon the first year analysis.
This forecast therefore serves as encouragement to the community to develop methodology and observations to constrain models of baryon feedback and enhance photometric redshift calibration at depths where spectroscopy is unrepresentative.
With tight priors on both these systematic terms, we forecast that LSST cosmic shear can deliver constraints on $S_8$ that are more than five times as constraining as existing cosmic shear surveys.

\end{abstract}

\maketitle
{\small \tableofcontents}
\vspace{1cm}
\section{Introduction}
\label{sec:intro}
%\linenumbers
%\rednote{Switching on the line numbers here has caused some footnotes to be pushed off the bottom of the page. See the non-line-numbered version for visible footnotes.}
Cosmology has entered a new era with the Vera C. Rubin Observatory and \Euclid satellite now building deep `full-sky' surveys\footnote{The \Euclid Wide Survey will image the `full' extragalactic sky outside the galactic and ecliptic planes, spanning roughly 14,000 square degrees \citep{euclid/mellier:2024}. 
The \Rubin Legacy Survey of Space and Time (LSST) will image the `full' Southern sky, spanning between 15,000 and 18,000 square degrees \citep{lochner/atal:2022}.}, exceeding the statistical power of all precursor imaging surveys.  
The study of weak gravitational lensing by large-scale structure, known as `cosmic shear', is a primary science driver for both \Euclid and LSST and the focus of this study. 
Cosmic shear measures the observed lensing-induced correlation between galaxy shapes as a function of redshift, mapping both the growth of large-scale structures and the expansion history of the Universe \citep[see][for example and references therein]{huterer:2023}.

The most recent precursor lensing surveys are close to completion: Dark Energy Survey \citep[DES,][]{bechtol/etal:2025}, Dark Energy Camera All Data Everywhere \citep[DECADE,][]{anbajagane/etal:2025}, Hyper Suprime-Cam \citep[HSC,][]{hiroaki/etal:2022}, Kilo-Degree Survey \citep[KiDS,][]{wright/etal:2024}. 
The primary weak lensing constraints made by these surveys are on the parameter $S_8\equiv\sigma_8\sqrt{\Omega_{\rm m}/0.3}$, which combines the matter density parameter, $\Omega_{\rm m}$, and the density fluctuation amplitude parameter $\sigma_8$\footnote{The parameter $\sigma_8$ is defined as the linear-theory standard deviation of matter density fluctuations in spheres of radius $8 h^{-1}{\rm Mpc}$ at $z=0$.}.
While these surveys find consistent percent-level measurements of these parameters \citep[see][hereafter DK23, and references therein]{DESY6/etal:2026,wright-cosmo/etal:2025,des/kids:2023}, there has been a degree of tension between them and equivalent CMB-derived measurements, in the context of $\Lambda$CDM \citep[see for example][]{divalentino/etal:2021}.
Efforts to understand differences have shone a spotlight on the various sources of astrophysical and data-related systematics in weak lensing studies \citep{mandelbaum:2018}. 
By taking a conservative approach to modelling these uncertainties, the precursor cosmic shear surveys are already limited from realising their full statistical power \citepalias{des/kids:2023}, a potentially problematic situation for \Euclid and \Rubin. 

A key missing ingredient from early \Euclid and LSST forecasts in \citet{Laureijs/etal:2011} and \citet[][hereafter DESC-SRD\footnote{DESC-SRD: the Dark Energy Science Collaboration - Science Requirements Document.}]{DESCSRD/etal:2018}, was the significant uncertainty in the theoretical modelling of the total matter distribution on small physical scales ($k>0.1h \, {\rm Mpc}^{-1}$). 
Using gravity-only $N$-body simulations, the non-linear matter distribution can be accurately determined either directly via an emulator \citep{heitmann/etal:2014,euclidemv2/etal:2021,derose/etal:2023,chen/etal:2025}, or indirectly using a simulation-calibrated halo model  \citep{takahashi/etal:2012,mead/etal:2015}. 
What these studies neglect, however, is the power that non-gravitional forces, with the dominant effect coming from Active Galactic Nuclei (AGN) in addition to supernovae and star formation processes, have to locally re-distribute both baryonic and dark matter \citep{white/etal:2004,semboloni/etal:2011,vandaalan/etal:2011}. 
Known as `baryon feedback', the significance of this astrophysical effect can be estimated using hydrodynamical simulations \citep{schaye/etal:2010,vandaalen/etal:2020,schaye/etal:2023,salcido/etal:2023,pakmor/etal:2023,mccarthy/etal:2023,martinalvarez/etal:2025}, analytical halo models \citep{debackere/etal:2020,mead/etal:2021}, or a process called `baryonification' whereby halo profiles in gravity-only simulations are modified to replicate different feedback scenarios \citep{schneider/teyssier:2015,arico/etal:2021,anbajagane/etal:2024}.   

Updated forecasts for \Euclid and LSST have assessed the impact of baryon feedback, quantifying the challenges related to this astrophysical systematic in a scenario where all other data-quality requirements are met.
\citet{martinelli/etal:2021} conclude that even for a `pessimistic' \Euclid analysis that removes projected small-scale cosmic shear information ($\ell_{\rm max} = 1500$), halving the nominal \Euclid dark energy figure of merit, the bias introduced from unmodelled baryon feedback is potentially significant enough to lead to a false detection of a time-varying dark energy equation of state. 
\citet{semboloni/etal:2013, huang/etal:2019,spuriomancini/bose:2023,boruah/etal:2024} demonstrate how removing small scale cosmic shear data and marginalising over a set of nuisance parameters to model the baryon feedback can remove this bias to the recovered cosmological parameters, at the cost of a loss in precision.
\citet{huang/etal:2019} find that baryon feedback modelling roughly doubles the LSST errors on the dark energy parameters.  
All studies concluded that the original forecast choice to mitigate baryon feedback through scale cuts alone, with $\ell_{\rm max} = 3000$ for LSST or $\ell_{\rm max} = 5000$ for \Euclid, was not a viable approach.

Recent results from the BAHAMAS, ANTILLES, FLAMINGO and XFABLE suites of hydrodynamical simulations\footnote{BAHAMAS was one of the first large-volume cosmological hydrodynamical simulations explicitly calibrated to reproduce key observations, with FLAMINGO its successor with much larger volumes, higher statistical power and explicit neutrinos modelling for precision cosmology. ANTILLES extends the BAHAMAS/FLAMINGO framework to explore a wider range of baryon feedback suppression and cosmological parameter variations used for emulator and inference applications. XFABLE tests stronger, more spatially extended AGN feedback to study its impact on the gas distribution and matter clustering.} \citep{mccarthy/etal:2017,salcido/etal:2023,schaye/etal:2023,bigwood/etal:2025a} serve to complicate the situation further.
These simulations are the first to reproduce the observed hot gas fraction in groups and clusters, the galaxy stellar mass function and the distribution of galaxy sizes in large cosmological volumes. 
They predict that baryon feedback could suppress the clustering of matter on physical scales as large as $\sim 60 {\rm Mpc}$ ($k \sim 0.15 h \, {\rm Mpc}^{-1}$), which is also predicted by halo models constrained by observed cluster gas fractions \citep{debackere/etal:2020}. 
In comparison, existing forecasts have primarily considered scenarios where baryon feedback only becomes significant on scales $k \gtrsim 1 h \, {\rm Mpc}^{-1}$.  
Observational evidence supports these more extreme models in the form of cross-correlation studies of weak lensing with either measurements of the thermal Sunyaev-Zel'dovich (SZ) effect \citep{amodeo/etal:2021,troester/etal:2022,pandey/etal:2023,mccarthy/etal:2023,to/etal:2024,dalal/etal:2025,laposta/etal:2025}, or the diffuse X-ray background \citep{ferreira/etal:2024}, along with joint analyses of lensing with the kinetic-SZ effect \citep{bigwood/etal:2024,mccarthy/etal:2024,hadzhiyska/etal:2025,sunseri/etal:2025,siegel/etal:2026}. Strong baryon feedback is also supported by studies of the thermal-SZ \citep{raghunathan/etal:2026,efstathiou/etal:2025,reichardt/etal:2021,bolliet/etal:2018} and fast radio bursts \citep{reischke/hagstotz:2025}. \citet{mcdonald/etal:2026} investigated the X-ray-weak lensing cross-correlation with FLAMINGO, however, and found no evidence for stronger feedback when comparing to observations, and furthermore they showed there is a near complete degeneracy between feedback and cosmology for this measurement.
These results motivate re-establishing the cosmic shear analysis choices for the next generation of imaging surveys. 
We note that while we analyse a mock Legacy Survey of Space and Time (LSST) for the first year of {\it Rubin}, our findings can be broadly applied to the \Euclid Wide Survey given the similarities in survey area and effective depth.

In this analysis we present cosmic shear forecasts for the first year (Y1) and final tenth year (Y10) of LSST, within a flat-$\Lambda$CDM cosmological model. 
Our main focus is the impact of baryonic feedback, as described above, but we also consider the interplay of our fiducial baryonic feedback model with a set of other effects: the influence of samplers, priors and the chosen parameter space; the effect of adopting two different models to account for the intrinsic alignment of galaxies \citep[see][and references therein]{chisari/etal:2025}; and the possibility that Y10 will not reach its required photometric redshift accuracy. The evolution from KiDS-1000 to KiDS-Legacy illustrates that redshift calibration remains one of the dominant systematic challenges in weak-lensing cosmology and the impact of improved spectroscopic calibration on $S_8$ constraints highlights the critical inclusion of spectroscopic datasets in maximising the cosmological return from next-generation surveys. With current techniques and instrumentation, the uncertainty on the redshift distribution of the Y10 sample will degrade relative to the Y1 sample, owing to the lack of a representative deep spectroscopic survey.  Such a sample exists for Y1 depths \citep{stanford/etal:2021}, allowing for an accurate and precise calibration of redshift distributions via machine learning methods \citep[see][and references therein]{newman/gruen:2022}.  The alternative calibration route for deep imaging surveys relies on clustering measurements between photometric and spectroscopic galaxy samples, where the spectroscopy spans the full redshift range, even if it is incomplete in colour-colour space \citep{newman:2008}.  Unfortunately this promising approach has only realised a redshift calibration precision that is more than an order of magnitude larger than the \citetalias{DESCSRD/etal:2018} Y10 requirement, limited by astrophysical uncertainty related to magnification and non-linear galaxy bias \citep{hildebrandt/etal:2021,gatti/etal:2022,wright-redshifts/etal:2025}.  We forecast how this level of degradation in the uncertainty on the redshift distribution will impact the Y10 constraints. 

Our approach differs from many previous LSST and \Euclid studies primarily because they forecast the $w_{\rm 0}w_{\rm a}$CDM model, constraining a time varying dark energy equation of state. 
Given that it has been argued that the hints of tension currently seen between independent probes for the $\Lambda$CDM model could be attributed to unmodelled systematic errors \citep{amon/efstathiou:2022,preston/etal:2023,elbers/etal:2024}, we seek to first define an analysis framework that can recover unbiased $\Lambda$CDM constraints from a realistic mock of LSST.
We argue that only if tensions and systematics in this subspace are fully understood is it safe to combine cosmic shear with the other data sets needed to obtain our target percent level constraints on the $w_{\rm 0}w_{\rm a}$CDM cosmological model.  As such, we focus on a $\Lambda$CDM forecast and only include $w_{\rm 0}w_{\rm a}$CDM forecasts in Section~\ref{sec:w0wa} as a means to compare our findings to other studies.

We have chosen to consider cosmic shear alone in this study, in contrast to creating a joint `3$\times$2pt' forecast which combines cosmic shear with galaxy clustering and galaxy-galaxy lensing \citep{heymans/etal:2021,DESY6/etal:2026,porredon/etal:2026}.  In the case where the set of adopted systematics models and priors provide an accurate representation of the underlying truth, a `3$\times$2pt' analysis will certainly help to self-calibrate the systematics that limit cosmic shear.  In the more realistic scenario, however, where the models are only approximate, the combination of model bias alongside the known interplay between the different systematic parameters can lead to errors in the joint constraints \citep{fischbacher/etal:2023, samuroff/etal:2024, leonard/etal:2024},  It is therefore important that each probe is explored and understood individually before moving forward with a joint analysis.

This paper is organised as follows: we introduce the cosmic shear analysis framework in Section~\ref{sec:methodology}, discussing both data-related and astrophysical systematic models along with our metric of success for the recovery of the input cosmological parameters. 
Section~\ref{sec:results} presents our $\Lambda$CDM forecasts, determining the rationale for our fiducial analysis of LSST-Y1 and exploring the robustness of that analysis to variations in the baryon feedback strength, survey depth and the intrinsic alignment model. 
In this forecast we sample the full likelihood instead of taking the more traditional Fisher matrix approach \citep{wolz/etal:2012}.  
For our $w_{\rm 0}w_{\rm a}$CDM forecasts in Section~\ref{sec:w0wa}, we explore weak lensing forecasts in combination with Gaussian priors based on the current generation of external data sets, minimising the computational expense of a full multi-probe forecast. 
We conclude in Section~\ref{sec:conclusions}. 
The main body of the paper is supplemented by a series of appendices: Appendix~\ref{app:theory} reviews the cosmic shear theory and the emulator used to rapidly calculate likelihoods; Appendix~\ref{app:mock_ingredients} discusses the creation of mock LSST cosmic shear observations;  Appendix~\ref{app:parametersandpriors} motivates our choice of cosmological parameter space and priors; Appendix~\ref{app:sampler} considers three different sampling algorithms and discusses our reasons for adopting \nautilus\footnote{\nautilus: \href{https://nautilus-sampler.readthedocs.io/}{nautilus-sampler.readthedocs.io/}} \citep{Naut_Lange23} for our analysis; Appendix~\ref{app:truth_params} details our approach to mitigate projection bias in our analysis; Appendix~\ref{app:extras} tabulates our key results along with additional figures of our constraints in extended parameter spaces.

\section{Methodology}
\label{sec:methodology}
We create mock cosmic shear observables for LSST-Y1 and Y10 using a five-bin tomographic convergence power spectrum $C_\ell$, implemented within \cosmosis \citep[see][and Appendix~\ref{app:theory}]{zuntz/etal:2015}, with the tomographic bins defined in \citetalias{DESCSRD/etal:2018}. 
Our ingredients are listed in Table~\ref{tab:mock_ingredients}, where we adopt the \citetalias{DESCSRD/etal:2018} parameters for LSST depth, area and expected data quality in terms of shear and redshift precision (for more detail see Appendix~\ref{app:mock_ingredients}). 

In Table~\ref{tab:priors} we list the parameters of the `truth' input cosmological model and astrophysical systematic models, compared to the set of priors that we adopt in our likelihood analysis.
The rationale for our choice of input cosmological parameters and priors is presented in Appendix~\ref{app:parametersandpriors}. 
For reference, Table~\ref{tab:priors} also includes the \citetalias{DESCSRD/etal:2018} adopted priors which in most cases are more informative. 
In this section we discuss how we model data-related systematics, baryon feedback and intrinsic galaxy alignment.

\begin{table*}
\centering                                      
\begin{tabular}{llcl}          
\toprule
Mock Ingredient     & Information  & \multicolumn{2}{l}{Reference} \\  
\midrule
Linear matter power spectrum & \camb & \multicolumn{2}{l}{\citet{lewis/etal:2000,howlett/etal:2012}} \\
Gravity-only non-linear power spectrum & \hmcode with no feedback & \multicolumn{2}{l}{\citet{mead/etal:2021}} \\
Baryon Feedback Suppression & FLAMINGO `L1$\_$m9' & \multicolumn{2}{l}{\citet{schaye/etal:2023}} \\
Intrinsic Alignment model & Non-linear linear alignment (NLA-z) & \multicolumn{2}{l}{\citet{bridle/king:2007,joachimi/etal:2011}} \\
\midrule
Cosmic Shear Statistic & Convergence Power Spectrum $C_\ell$  & \multicolumn{2}{l}{\citet{kilbinger/etal:2017}, equation 43(ii)} \\
Number of $\ell$ bins & 24 log-spaced & &\\
Mock $\ell$-range & $\ell=[20,5000]$ &  &\\
\midrule
Ellipticity Variance & $\sigma_e=0.26$ &  \citetalias{DESCSRD/etal:2018}: & appendix D2.1  \\
Masked area fraction & $f_{\rm mask} = 0.12$ & " & appendix F1\\
Number of Redshift Bins & 5, equal numbers in each bin & " & appendix D2.1  \\
Photometric Redshift Error & $p(z_{\rm p} | z_{\rm true}) = {\cal N}\left[\mu =z_{\rm true} ;\sigma = 0.05(1+z_{\rm true})\right]$ & " & appendix D2.1 \\
Known Photometric Redshift Outliers & 7\% uniformly distributed between $0<z<3$ & & \\
 \midrule
Y1 Area & $12,300 \, \mathrm{deg}^2$ & " & appendix C1 \\
Y1 Depth ($5\sigma$ point source) & ${\it i}<24.13$ & " & appendix C1 \\
Y1 Number density & $n_{\rm eff} = 11.112 \times (1-f_{\rm mask} ) = 9.78 \, {\rm arcmin}^{-2}$ & " & table F1\\
Y1 Redshift Distribution & $n(z)  \propto z^2 \exp[-(z/0.13)^{0.78}]$ \,\, $\bar{z} = 0.85$ & " & figure F4  \\
\midrule
Y10 Area & $14,300 \, \mathrm{deg}^2$ & " & appendix C1 \\
Y10 Depth ($5\sigma$ point source) & ${\it i}<26.35$ & " & appendix C1 \\
Y10 Number density & $n_{\rm eff} = 27.737 \times (1-f_{\rm mask} ) = 24.4 \, {\rm arcmin}^{-2}$ & " &  table F1\\
Y10 Redshift Distribution & $n(z) \propto z^2 \exp[-(z/0.11)^{0.68}]$ \,\,$\bar{z}  = 1.05$ & " & figure F4 
\\ 
\bottomrule
\end{tabular}
\caption{A list of ingredients for mock cosmic shear observations of LSST-Y1 and Y10: the input theoretical models for the cosmological and contaminating signals; the cosmic shear statistic; the simulated data quality in terms of shear and redshift precision; Y1 and Y10 survey parameters.  References are provided in the final column and discussed in Appendices ~\ref{app:theory} and~\ref{app:mock_ingredients}. We note that the LSST survey footprint will remain under review throughout the survey's lifetime, with the final area expected to be closer to $18,000 \, \mathrm{deg}^2$, $\sim 25\%$ larger than the \citetalias{DESCSRD/etal:2018} survey parameters adopted in this analysis.  As we find that our LSST cosmic shear forecasts are systematics dominated, an increase in survey area would not impact our conclusions.
}
\label{tab:mock_ingredients}
\end{table*}

\begin{table*}
\centering                                      
\begin{tabular}{llllc}          
\toprule
 & Input & \citetalias{DESCSRD/etal:2018} & Fiducial Analysis & Informed? \\    
\midrule
\multicolumn{3}{l}{\bf Cosmological parameter priors:}\\  
Amplitude & $\sigma_8= 0.8$ & $\sigma_8: {\cal N}(0.831; 0.14)$ & $\sigma_8: \bb{0.39; 1.01}$ & \xmark \\
Hubble constant & $h= 0.715$ & $h: {\cal N}(0.6727; 0.063)$ & $h: \bb{0.65,0.78}$ & \cmark\\
Matter density & $\Omega_{\rm m}=0.3$ &$\Omega_{\rm m}:{\cal N}(0.3156; 0.2)$ &  $\Omega_{\rm m}:\bb{0.2; 0.46}$ & \xmark \\
Baryon density& $\omega_{\rm b}= 0.02233$ & $\Omega_{\rm b}: {\cal N}(0.0492; 0.006)$ & $\omega_{\rm b}: {\cal N}(0.02233,0.0004)$ & \cmark \\
Spectral index & $n_{\rm s}= 0.97$ & $n_{\rm s}: {\cal N}(0.9645;0.08)$ &  $n_{\rm s}: \bb{0.95,\,0.99}$ & \cmark\\
Neutrinos & $\Sigma m_\nu =  0.06$ & $\Sigma m_\nu = 0.00 {\rm eV}$ &  $\Sigma m_\nu = 0.06 {\rm eV}$ & \cmark\\
Dark Energy &$w_{\rm 0}=-1$ & $w_{\rm 0}: {\cal N}(-1.0;0.8)$ & $w_{\rm 0}=-1$ & \cmark\\
 & $w_{\rm a}=0$ & $w_{\rm a}: {\cal N}(0.0;2.0)$& $w_{\rm a}=0$ & \cmark\\
\midrule      
\multicolumn{3}{l}{\bf Astrophysical systematic models and priors:}\\                             
Intrinsic Alignments & NLA-$z$ & NLA+ &  NLA-z  & \\
 & $A_{\rm IA}=0.4$ & $A_{\rm IA}: {\cal N}(5;3.9)$, $\eta_{\rm IA}: {\cal N}(0;2.3)$ & $A_{\rm IA}: \bb{-0.6,1.2}$ & \xmark\\
 & $\eta_{\rm IA}=2.2$ &$\beta_{\rm IA}: {\cal N}(1;1.6)$, $\eta_{\rm z}: {\cal N}(0;0.8)$ &  $\eta_{\rm IA}: \bb{0.1,5}$ & \xmark\\
Non-linear Model & \hmcode & \halofit &  \hmcode & \\
Baryon Feedback &  FLAMINGO L1\textunderscore m9 &  $\ell < 3000$  &  Scale cuts: $K_{\rm max}=0.5$ & \\
& & &  $\Theta_{\rm AGN}:\bb{7.4,7.95}$ & \cmark \\
Neutrino Model & \hmcode & N/A  & \hmcode  \\
\midrule
\multicolumn{3}{l}{\bf Shear and Redshift Calibration:}\\                                
$\sigma_m^i$ & 0 & $|0.013(2z_i-z_{\rm max})/z_{\rm max}|$ &   $0.002(1+z)$ & \cmark \\
$\sigma_z^i$ & 0 & $0.002(1+z)$ &   $0.002(1+z)$ & \cmark \\
Correlated errors? & N/A & No & Yes & \\
\bottomrule
\end{tabular}
\caption{Cosmological parameters, astrophysical systematic models, catalogue properties and the calibration parameters: listing the input values to the mock observations and comparing our fiducial modelling and prior choices with \citetalias{DESCSRD/etal:2018}. 
The DESC-SRD used Gaussian priors, as this was a Fisher forecast, indicated as ${\cal N}(\mu;\sigma)$ with a mean $\mu$ and variance $\sigma^2$. 
In this analysis we have chosen top-hat priors with the limits indicated in square brackets, and the final column indicating whether these priors are informed, \cmark, or uninformed, \xmark $\,$ (see Appendix~\ref{app:parametersandpriors}). 
The listed cosmological parameters are: the density fluctuation amplitude parameter, $\sigma_8$; the Hubble constant, $h=H_0/(100\,{\rm km}\,{\rm s}^{-1} \, {\rm Mpc}^{-1})$; the matter density, $\Omega_{\rm m}$; the baryon density, $\omega_{\rm b}=\Omega_{\rm b}h^2$; the scalar spectral index, $n_{\rm s}$; the sum of the neutrino masses, $\Sigma m_\nu$.} 
\label{tab:priors}           
\end{table*}

\subsection{Data-related systematics}
\label{sec:datasys}
Cosmic shear measures the lensing induced by the projected mass distribution along the line of sight. 
The signal is therefore highly sensitive to the mean redshift of the galaxy population. 
In this study, we limit our model of the redshift calibration uncertainty for each tomographic bin to a single parameter that measures the bias in the mean redshift, $\delta_{\rm z}^i$ such that $n^i_{\rm estimate}(z) = n^i_{\rm true}(z+\delta_{\rm z}^i) $. 
We take a different approach from \citetalias{DESCSRD/etal:2018}: instead of calculating the required limit for $\delta_{\rm z}^i$ to ensure LSST achieves a particular figure of merit, we ask what is feasible given current instrumentation and methodology.   

At Y1 depth, the photometric galaxy sample is well-matched to the C3R2 spectroscopic survey, designed to provide galaxy spectroscopy for a sample that is complete in colour-space to $i\sim 24.5$ \citep[$10\sigma$ extended source,][]{stanford/etal:2021}. 
With C3R2, Y1 photometric redshifts can be calibrated using self-organising maps (SOM) and a two-step process utilising the Deep Drilling Fields to improve the calibration accuracy for the Wide survey \citep[see for example][]{buchs/etal:2019,wright/etal:2020,mccullough/etal:2024,yin/etal:2026}. 
\citet{stanford/etal:2021} demonstrate that the existing C3R2 sample can already deliver an accuracy of $\delta_{\rm z} \sim 0.002(1+z)$ for 92\% of the \Euclid sample, which is slightly deeper than Y1. 
As C3R2 continues to collect data, we adopt $\delta_{\rm z}^i \sim 0.002(1+z_i)$ for our fiducial Y1 forecast, which also matches the \citetalias{DESCSRD/etal:2018} requirement.  
We adopt a multivariate Gaussian prior\footnote{This can be implemented using the {\tt correlated\_priors} module within {\sc CosmoSIS} which uses a Cholesky decomposition of the covariance matrix between nuisance parameters to create multivariate parameters that follow normal distributions.} for the vector of redshift nuisance parameters, $\delta_{\rm z}^i$, in order to implement an assumed 20\% correlation in the redshift uncertainty between neighbouring bins and a 5\% correlation between non-neighbouring bin pairs, as motivated by \citet{hildebrandt/etal:2021,cordero/etal:2022}.

At Y10 depth, photometric redshift calibration will most likely be determined using cross-correlation techniques \citep{newman:2008}.
By measuring the angular clustering between the Y10 photometric sample and a luminous spectroscopic sample that spans the full redshift range, the redshift distribution of each tomographic bin can be estimated. 
For Stage-III surveys, the accuracy of this approach has been shown to be limited by uncertainty on the magnification and galaxy bias model for the spectroscopic sample. 
This results in a calibration uncertainty $\delta_{\rm z}^i \sim 0.015$ that is broadly independent of redshift \citep{hildebrandt/etal:2021,gatti/etal:2022, newman/gruen:2022}. 
In principle, the spectroscopic galaxy bias could be self-calibrated using a `$6 \times 2$pt' analysis, cross-correlating the shear with itself and the photometric and spectroscopic samples, in addition to auto- and cross-clustering of the two galaxy samples \citep{bernstein:2009}.  As this approach remains under development, however \citep[see the successes and challenges discussed in][]{johnston/etal:2024}, we choose to adopt $\delta_{\rm z}^i = 0.015$ for our Y10-like forecast, including
an optimistic scenario variant with $\delta_{\rm z}^i \sim 0.001(1+z_i)$. 

In this analysis we limit our modelling of redshift uncertainty to the standard $\delta_{\rm z}^i$ linear shift parameters, assuming that the outlier population is known accurately.  This approach is sufficient for current surveys \citep{stolzner/etal:2021,amon/etal:2022}, but for LSST and \Euclid, additional $n(z)$ nuisance parameters will be required to ensure the accurate recovery of cosmological constraints \citep{boruah/etal:2024,mill/etal:2025,DESY6/etal:2026}.  The community has yet to converge on an optimal approach to flexibly model the $n(z)$ \citep{stolzner/etal:2021,cordero/etal:2022, myles/etal:2023, ruiz-zapatero/etal:2023,zhang/etal:2023b, reischke:2024,bernstein/etal:2025,dassignies/etal:2026}.

For each tomographic bin $i$, we marginalise over our uncertainty on the calibration of the measured shear, $\sigma_{\rm m}^i$, following \citet{li/etal:2023}, who simulate a range of different galaxy morphology distributions to quantify the shear calibration uncertainty arising from imperfect image simulations.  
Assuming that numerous image simulations can be analysed to minimise statistical noise on the shear calibration measurement, $m$, they find a limiting systematic uncertainty at the level $\sigma_{\rm m} \sim 0.002(1+z)$.  This primarily arises from the sensitivity of $m$ to changes in the distribution of intrinsic galaxy ellipticities. 
We choose to adopt this value for our assumed shear calibration uncertainty, including $\sigma_{\rm m}^i$ as an additional error to the analytical covariance described in Appendix~\ref{app:mock_ingredients} \citep[see equation 37][for details]{joachimi/etal:2021}.   Motivated by the fact that the properties of the galaxies that give rise to this shear calibration uncertainty vary slowly between tomographic bins, we include 100\% correlation in $\sigma_{\rm m}^i$ across the $i$ tomographic bins. This approach is equivalent to marginalising over a single free $m$ parameter and scaling it by fixed $\sigma_{\rm m}^i$ values, such that all the effective $m$ values are 100\% correlated \citep{asgari/etal:2021}.

Our choice is optimistic compared to the shear calibration requirements set by \citetalias{DESCSRD/etal:2018}, suggesting that shear measurement will not be the limiting systematic for LSST cosmic shear.  It is also pessimistic compared to the accuracy forecasted for the DESC-adoped \textit{Metadetection} technique \citep{sheldon/etal:2020}.  Using a deconvolution approach to account for noise-, model-, and detection-related bias, combined with deep data from the LSST Deep Drilling Fields, this method has been shown to deliver a shear accuracy that is better than $\sigma_\mathrm{m} < 0.001$ \citep{zhang/etal:2023,sheldon/2023}.  We note that this high level of accuracy is expected to degrade, however, when redshift-dependent blending is included in the forecast analysis.

\subsection{Baryon Feedback}
\subsubsection{Building mock data using FLAMINGO}
We include the suppression due to baryon feedback in our mock data vector, using the FLAMINGO hydrodynamical simulations \citep{schaye/etal:2023}. 
In this suite the four parameters governing the sub-grid physics of the stellar and AGN feedback mechanisms are pre-defined to ensure the observed cluster gas fraction and the stellar mass function can be re-produced within the measurement uncertainty \citep{kugel/etal:2023}.  Assuming that the observational constraints on the hot gas content of galaxy clusters inferred from X-ray observations are not subject to selection effects or unrecognised biases, the use of FLAMINGO provides a realistic range of input feedback models for our study.  

\citet{schaye/etal:2023} simulate eleven different observationally-consistent combinations of the sub-grid parameter set, with corresponding gravity-only simulations.
For our fiducial mock we adopt the fiducial FLAMINGO simulation `L1$\_$m9', a $1~\rm{Gpc}$ intermediate resolution simulation which provides the closest match to the calibration observables (stellar mass function and cluster gas fraction) whilst also recovering the non-calibration observables (e.g. cosmic star formation history; galaxy size and metalicity; X-ray and SZ cluster scaling relations).  
Using measurements of the three-dimensional matter power spectrum, $P_{\rm mm}(k,z)$ \citep{schaller/etal:2024}, we take the ratio between the `L1$\_$m9' hydrodynamical simulation and the corresponding gravity-only simulation as our feedback suppression model.  
This ratio is then applied to the gravity-only non-linear matter power spectrum, generated using \hmcode, to mitigate the noise in the FLAMINGO realisations. 
We then produce predictions for the simulated $C_\ell$ cosmic shear power spectrum (see Appendix~\ref{app:theory}). 
Analysing the wide range of alternative FLAMINGO simulations, which also reliably reproduce the range of different observations, we find that the `fgas+2$\sigma$' and `Jet\_fgas-4$\sigma$'\footnote{The FLAMINGO jet model is additionally interesting as the baryonic suppression curve has a different shape compared to the other FLAMINGO simulations, (Cosmo-)OWLS and BAHAMAS, yielding a stronger suppression on large physical scales for a fixed cluster gas fraction \citep{schaye/etal:2023}.} scenarios induce the weakest and strongest suppression respectively, in terms of $C_\ell$ \citep[see table 1 of][for details of the sub-grid physics of each scenario]{schaye/etal:2023}. We use the `fgas+2$\sigma$' and `Jet\_fgas-4$\sigma$ FLAMINGO models as input to our mocks when testing the sensitivity of our analysis to weak and strong feedback scenarios.

\begin{figure*}
    \centering
    \includegraphics[width=1.0\textwidth]{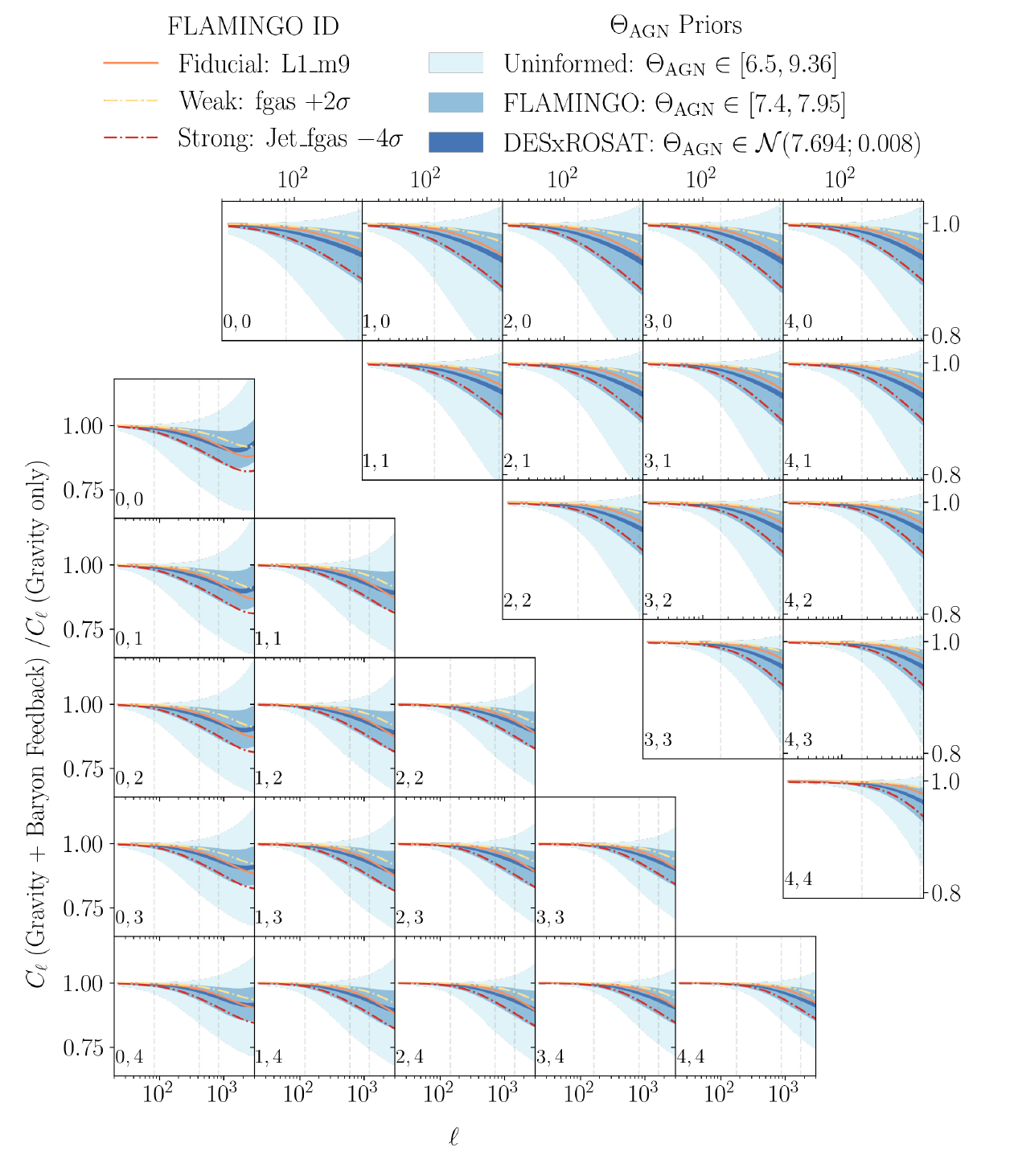}
    \caption{Cosmic shear power spectra, $C_{\ell}^{ij}$: the ratio between models including baryon feedback and the gravity-only scenario, for the Y1 redshift distributions and assuming the cosmology detailed in Table~\ref{tab:priors}.
    Each subplot corresponds to a different combination of tomographic bins $i$ and $j$ with the auto-correlations shown on the diagonal.
    The lower-left panel shows the uncertainty in baryon feedback suppression over the \citetalias {DESCSRD/etal:2018} analysis range $20<\ell<3000$. 
    The upper-right panel zooms into the same data, but over our fiducial analysis range with $K_{\rm max}=0.5$.
    We model baryon feedback using the FLAMINGO simulations \citep{schaye/etal:2023}, using their `L1\_m9' simulation as our fiducial case (orange), alongside their weakest (`fgas+2$\sigma$' in yellow) and strongest (`Jet\_fgas-4$\sigma$' in red) scenarios. 
    The shaded regions correspond to the three different $\Theta_{\rm AGN}$ prior ranges considered in our \hmcode analysis.  We compare an uninformative prior (light-blue), alongside two informative priors: one spans the same region as the FLAMINGO simulations at low-$\ell$ (mid-blue), and one adopts tight constraints from external DESxROSAT data (dark-blue). The vertical dashed grey lines correspond to $K_{\rm max}=[0.1, 0.5, 1.0]$ in the lower-left panel, and $K_{\rm max}=[0.1, 0.5]$ in the upper-right panel. }
    \label{fig:FLAMINGO}
\end{figure*}

Figure~\ref{fig:FLAMINGO} shows the suppression of the cosmic shear power spectrum, $C_{\ell}^{\rm with \,baryons}/C_{\ell}^{\rm gravity \, only}$, for our fiducial mock and the weakest and strongest extremes from the FLAMINGO suite.
Each box corresponds to the cosmic shear power spectrum suppression for a given pair of tomographic bins, with the auto-correlation shown on the diagonal.
The lower-left corner plot shows the \citetalias{DESCSRD/etal:2018}-adopted range of scales with $20 \leq \ell \leq 3000$, where the feedback suppression can be as strong as $\sim 20$\%.

\label{sec:feedback}

\subsubsection{Scale Cuts}
In this analysis we combine two approaches to mitigate our uncertainty on the impact of baryon feedback. 
We introduce scale cuts to remove data on highly non-linear scales and include nuisance parameters to model our uncertainty on the impact of baryon feedback on the remaining scales. These two approaches follow similar methodology applied in previous weak lensing analyses: in the DES-Y6 analysis scale cuts were applied to mitigate the impact of baryon feedback \citep{sanchez-cid/etal:2026}, whereas HSC-Y3 and KiDS-Legacy chose to include smaller scales and marginalise over their uncertainty on the strength of the baryon feedback using \hmcode \citep{dalal/etal:2023,wright-cosmo/etal:2025}.

Starting with the scale cut definition, Figure~\ref{fig:FLAMINGO} demonstrates that any optimised strategy will require redshift dependence, applying more conservative $\ell$-cuts to the lower redshift bins.
We therefore define a set of scale cuts, denoted\footnote{We use the notation $K_{\rm max}$ to signify the relationship to a selection in $k-$space.   
If the galaxies in the two tomographic bins of interest were all located at a single redshift, then our defined $\ell_{\rm max}$ cut would be equivalent to a $k$-cut at $k=K_{\rm max}$.
With the reality of broad redshift distributions, however, the projected cosmic shear signal at each $\ell$-scale contains input from the underlying matter power spectrum over a wide range of $k$-scales.
As such the $K_{\rm max}$ selection is only an indication of the $k$-scale contribution.} $K_{\rm max}=[0.1, 0.3, 0.4, 0.5, 0.6, 0.7, 1.0]$, where data in the tomographic bin combination $ij$ is included when $\ell<\ell_{\rm max}^{ij}$, with 
\begin{equation}
\ell_{\rm max}^{ij} = K_{\rm max} \chi^{ij}.
\label{eqn:Kmax-lmax}
\end{equation}
Here the minimum co-moving distance $\chi^{ij}=\rm{min}\left[\chi(\bar{z_i}), \chi(\bar{z_j})\right]$, where $\bar{z_i}$ is the median redshift of the tomographic redshift bin $i$.  For alternative methods to define scale cuts see \citet{gu/etal:2025,piccirilli/etal:2025,zanoletti/leonard:2025,truttero/etal:2025,robertson/hall:2026}.
Our $K_{\rm max}=[0.1, 0.5, 1]$ selection is shown in the lower-left corner plot of Figure~\ref{fig:FLAMINGO} as the dashed grey vertical lines, with the corresponding $\ell_{\rm max}$ for each tomographic bin tabulated in Appendix~\ref{app:extrastab}. 
In the top-right corner plot we zoom into the $K_{\rm max}=0.5$ limited range of scales that are used for our fiducial analysis configuration (see Section~\ref{sec:fiducial_results}). Note that we keep the large-scale cut at $\ell=20$, throughout the analysis following the \citetalias{DESCSRD/etal:2018}.

\subsubsection{Fiducial analysis with \hmcode}
For our fiducial analysis we adopt \hmcode to model the non-linear matter power spectrum \citep{mead/etal:2021}, including the free parameter $\Theta_{\rm AGN}$ to modulate the strength of the baryon feedback suppression. $\Theta_{\rm AGN}$ is an internal parameter of \hmcode, related to the AGN heating parameter in the BAHAMAS simulations as $\Theta_{\rm AGN}=\log_{10}(\Delta T_{\rm heat}/{\rm K})$. Six parameters simultaneously vary within \hmcode's halo model in order to recover the BAHAMAS-predicted suppression for a range of $\Delta T_{\rm heat}$ at percent-level accuracy. We note that any constraint on $\Theta_\mathrm{AGN}$ has no physical meaning about the temperature of the AGN and that the functional form of the baryon feedback suppression predicted by \hmcode does not match the FLAMINGO results precisely, as seen in Figure~\ref{fig:FLAMINGO} (see~\ref{sec:intro} for discussion on differences between simulations). 
There are a range of observational constraints on this parameterisation of baryon feedback, with \citet{troester/etal:2022} using a cross-correlation analysis of weak lensing and the thermal SZ effect finding $\Theta_{\rm AGN} = 7.96^{+0.2}_{-0.48}$ and \citet{ferreira/etal:2024} using the cross-correlation between cosmic shear from DES-Y3 with ROSAT X-ray data (hereafter DESxROSAT) to constrain $\Theta_{\rm AGN}=7.998\pm0.008$.
These studies highlight the opportunity to use external data to set informative priors on baryon feedback models in cosmic shear. 
Alternatively, we can look to simulations to define an informed prior based on the $\Theta_{\rm AGN}$ range (or any other baryon feedback parameterisation) for a given suite of simulations. 

In this analysis we consider three different priors for $\Theta_{\rm AGN}$, shown as shaded regions in Figure~\ref{fig:FLAMINGO}, along with a gravity-only analysis where we use \hmcode with the no feedback option.  For an `uninformed' prior\footnote{We note that the lower-limit of this prior is only uninformative in a cosmic shear analysis of the full $\ell$-range shown in the lower right corner plot of Figure~\ref{fig:FLAMINGO}.} we use a top hat $\Theta_{\rm AGN} \in [6.5, 9.36]$ with the lower bound defined using the $1 \sigma$ lower limit of the gravity-only fit, and the upper bound defined as $+7 \sigma$ from the \citet{troester/etal:2022} constraint.
The FLAMINGO-informed prior is defined by the values of $\Theta_{\rm AGN} \in [7.4, 7.95]$ that span the full FLAMINGO range, encompassing the `strong' and `weak' versions of our mock data vector shown as the dashed lines in Figure~\ref{fig:FLAMINGO}. 
Finally, for the DESxROSAT observation-informed Gaussian prior with $\Theta_{\rm AGN} \in {\cal N}(7.694; 0.008)$, the width is taken from \citet{ferreira/etal:2024}, with the central value chosen to be the best-fit value to the fiducial mock when all other parameters are fixed to the truth.

For our fiducial \hmcode analysis we choose to use the FLAMINGO-informed prior, presenting results for the other two priors to quantify our sensitivity to this choice.  The hydrodynamical simulations are unable to recover key observables outwith this range making our uninformative prior arguably over-conservative.  In contrast the DESxROSAT constraints are likely over-optimistic since they derive from an analysis that assumes a fixed cosmology and metallicity value for the intra-cluster medium without accounting for the uncertain contribution to the signal from unresolved clustered X-ray AGN.  The use of external observations like DESxROSAT, will likely be central to exploiting future cosmic shear observations, but given the complex astrophysical modelling that is required to interpret any `mass-gas' cross-correlation data \citep[see for example][]{mead/etal:2020}, the level of confidence in this type of constraint will likely be debated. 

There are many alternative approaches to using \hmcode to model baryon feedback including: calibrated baryonification models \citep[see for example][]{bigwood/etal:2025b}, updated hydrodynamical simulation-informed halo models \citep[see the][`resummation model']{vandaalen/etal:2026}, along with emulators, some of which are created using machine learning techniques \citep{kammerer/etal:2025,lin/etal:2025} or an effective field theory inspired approach \citep{derose/etal:2025}.  In this analysis, we choose an emulator-based alternative to \hmcode, modelling the impact of baryons using the \spk model\footnote{\spk: \href{https://github.com/jemme07/pyspk}{github.com/jemme07/pyspk}} \citep{salcido/etal:2023}. 
\spk is trained on the ANTILLES suite of 400 hydrodynamical simulations that cover a wide range of stellar and AGN efficiencies, providing a diverse and flexible emulator model. 
One of the key strengths of this model, compared to \hmcode, is that the free parameters are tied to the variation of the baryon fraction as a function of halo mass $f_{\rm b} (M_{\rm halo})$. 
In principle, this function can be directly constrained with observations, in contrast to sub-grid feedback parameters such as $\Theta_{\rm AGN}$. \citet{salcido/etal:2023} show that \spk can recover the ANTILLES matter power spectra to a precision of 2\% to $k<10h\,{\rm Mpc}^{-1}$.  

\subsubsection{Alternative analysis with \spk}
We use the most flexible redshift-dependent double power-law version of \spk to marginalise over our uncertainty on baryon feedback.  
This \spk version models $f_{\rm b} (M_{\rm halo})$ using four free parameters: the amplitude, $\varepsilon$; two power-law slopes at low mass, $\alpha$, and high mass, $\beta$; a redshift dependence, $\gamma$.
Here we mirror the approach of our fiducial \hmcode analysis by setting informative top hat priors on these nuisance parameters, defined to span the range of the FLAMINGO simulations: $\varepsilon \in [0.24, 0.35]$, $\alpha \in [-0.12, 0.34]$, $\beta\in [-0.74, 0.77]$, $\gamma \in [1.02, 1.20]$.  These parameter ranges are within the ANTILLES training sample, with the exception of the redshift dependence parameter $\gamma$. Compared to ANTILLES, the FLAMINGO simulations display stronger redshift evolution.

One important caveat to note is that our analysis is underpinned by three related hydrodynamical simulations.  With the exception of the jet model used in our `strong' baryon mock, BAHAMAS (used to calibrate \hmcode), ANTILLES (used to train \spk) and FLAMINGO (used to create our mocks and baryon feedback priors) derive from a similar family of models.  The agreement found between these suites can therefore not be viewed as independent confirmation that the power spectra response considered in this analysis is broad enough to cover the full impact and scale-dependence of baryon feedback on cosmic shear.  

\subsection{Intrinsic Galaxy Alignment}
Galaxies evolve within dark matter haloes that interact with the large-scale matter distribution via gravity.
Through these interactions, the underlying tidal field coherently aligns neighbouring galaxies, which we refer to as intrinsic galaxy alignment.
In practice, the observed correlations between galaxy orientations combine the lensing effect of matter with the contribution from intrinsic alignments, which therefore needs to be accounted for when modelling the cosmic shear signal. 
Direct observations of the intrinsic alignment of galaxies are wide-ranging \citep[see][and references therein]{johnston/etal:2021}.
Luminous red galaxies are found to exhibit a strong alignment with their local density field \citep[\eg][]{mandelbaum/etal:2006}.  
Satellite galaxies align radially towards their central galaxy \citep[\eg][]{georgiou/etal:2019}.  
For blue field galaxies, however, an alignment signal has yet to be detected \citep[\eg][]{siegel/etal:2025}.  Unfortunately, these galaxy type-specific constraints from the literature cannot be directly incorporated in our analysis as, for a magnitude-limited cosmic shear survey like LSST, each tomographic bin will contain a different fraction of red-blue and central-satellite galaxies.  

In this analysis we choose to adopt the redshift-dependent non-linear linear alignment model \citep[NLA-z, Equation~\ref{eqn:NLA-z},][]{bridle/king:2007, joachimi/etal:2011} following the recommendation of \citet{fortuna/etal:2021}.  They model the intrinsic alignment of an evolving galaxy sample using a halo model to combine constraints on galaxy type-specific alignment to predict the overall IA contamination for a Stage IV survey.  By running a cosmological analysis of halo model IA-contaminated cosmic shear signals, whilst adopting the NLA-z model in their analysis framework, they determine a set of effective NLA-z parameters.  The biggest uncertainty in the IA halo model arises from the uncertainty in the luminosity dependence of the red galaxy alignment. \citet{fortuna/etal:2021} present parameters using a simple power-law in luminosity ($A_{\rm IA}=0.16 \pm 0.02$ and $\eta_{\rm IA}=2.91 \pm 0.70$) and a broken power-law ($A_{\rm IA}=0.42 \pm 0.02$ and $\eta_{\rm IA}=2.21 \pm 0.22$).
For a Stage IV cosmic shear analysis where the underlying truth is given by an IA halo model with the alignment amplitudes determined from observations, they conclude that the adoption of an NLA-z model would not introduce a significant bias in the cosmological parameter constraints. 
We therefore use the two-parameter NLA-z model in our fiducial analysis with uninformative top-hat priors $A_{\rm IA}:\bb{-0.6,1.2}$ and $\eta_{\rm IA} = \bb{0.1,5.0}$. 
The edges of these priors represent $\pm 7 \mathcal{D}$ about the mid-point of the simple and broken power-law constraints from \citet{fortuna/etal:2021}, with $\mathcal{D}$ defined as the difference between the two results.

It has been argued that the NLA-z intrinsic alignment model is ad hoc, and we therefore include an alternative analysis utilising the Tidal Alignment and Tidal Torquing model \citep[TATT, see equations 37-39 of][for the intrinsic alignment power spectra]{blazek/etal:2019}. 
This five-parameter TATT model combines a linear alignment model with a tidal torquing alignment mechanism and a density weighting term.  

\subsection{Posterior Summary Statistics and Projection Effects}
\label{sec:truthtest}
In this cosmic shear analysis we define our metric of success based on the accurate recovery of the most constrained cosmological parameter, $S_8$. 
There are several summary statistics that could be employed when quoting parameter constraints, with the maximum or mean marginal distribution most often reported alongside a corresponding credible interval. 
It is well known, however, that these two summary statistics do not always peak at the true parameter value, nor does the one-dimensional credible interval necessarily encompass it, as a result of projection effects \citepalias[see appendix C.3 of][and references therein]{des/kids:2023}.
A prior space that is asymmetric about the truth can lead to offsets between the peak of the marginalised posterior and the MAP (maximum a posteriori), when sampling over a multi-dimensional parameter space with non-linear degeneracies between constrained and unconstrained parameters.   

To date, studies have mainly circumvented this issue of `projection bias' by measuring relative offsets in the mean marginal between a `truth' analysis, where the analysis framework perfectly matches the input noise-free mock, and a series of robustness `test' analyses where different aspects of the framework are varied.  Success is declared when the `test' results match the `truth' results, rather than the input set of cosmological parameters. We adopt this approach with a key modification introduced by \citet{derose/etal:2022,blake/etal:2025}: as projection effects depend on the choice of priors, scale cuts and the baryon feedback strength, a bespoke `truth' analysis must be created to match each `test' case.  For example, when investigating the accurate recovery of cosmological parameters in the case of strong baryon feedback, our `test' case will use \hmcode or \spk to analyse a mock data vector created with the `Jet\_fgas-4$\sigma$' FLAMINGO model.  Our matching \hmcode `truth' case will analyse a mock data vector created with \hmcode where $\Theta_{\rm AGN}=7.931$. This value of $\Theta_{\rm AGN}$ is the best-fitting \hmcode model to the `Jet\_fgas-4$\sigma$' case, when all other parameters are fixed to the input values listed in Table~\ref{tab:priors}.  Our matching \spk `truth' case will analyse a different but similar mock data vector created with \spk (see Appendix~\ref{app:truth_params} for details).  The mean marginal $S_8$ from the bespoke `truth' analysis determines the level of `projection bias' for each case.  The offset in $S_8$ relative to the matching `test' result then determines the accuracy of that analysis.

Our threshold for success is set at $\Delta S_8 = | S_8^{\rm truth} - S_8^{\rm test}| < 0.5 \sigma$ where $S_8^{\rm x}$ is the mean marginal estimate from the `truth' and `test' chains, with $\sigma$ defined as the 68\% credible interval estimated from the `test' chain.  
There are various reasonable choices for these criteria from highly restrictive requirements like $0.1\sigma$ in \citet{joachimi/etal:2021} to broader choices such as a joint 2D $(S_8, \Omega_{\rm m})$ plane \citet{krause/etal:2021}. Our choice is comparable to the latter since we define it with respect to the single parameter $S_8$ value.

\section{Results}\label{sec:results}
\begin{figure*}
    \centering
    \includegraphics[width=\textwidth]{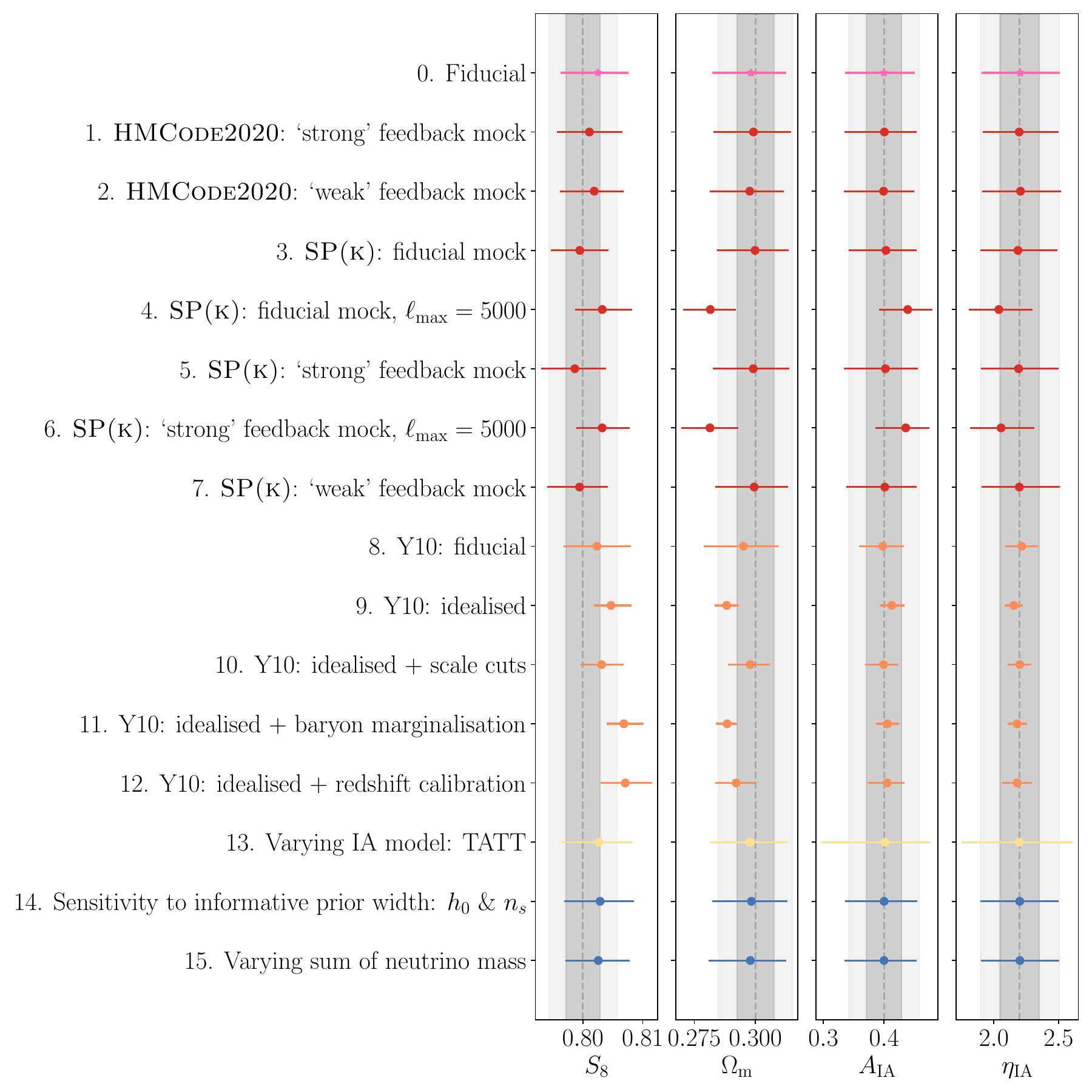}
    \caption{LSST $\Lambda$CDM cosmic shear forecasts for $S_8$, $\Omega_{\rm m}$ and the NLA-$z$ intrinsic alignment parameters $A_{\rm IA}$ and $\eta_{\rm IA}$, showing the mean marginal and 68\% credible interval.
    Our Y1 fiducial analysis configuration is shown at the top in pink, with each row that follows presenting a series of variant analyses as detailed in the text.  With the exception of the variants noting the use of a `weak' or `strong' feedback mock, we incorporate baryon feedback using the fiducial FLAMINGO model shown in Figure~\ref{fig:FLAMINGO}. Aside from variants noted as using scales up to an $\ell_{\rm max}=5000$, we apply a scale cut of $K_{\rm max}=0.5$.
    The dark grey region provides an indication of our target accuracy of $\pm 0.5\sigma$ uncertainty, about the `truth' (shown dashed), where $\sigma$ is the 68\% credible interval from our fiducial Y1 analysis.  Note that here we show $S_8 = S_8^{\rm input} + \Delta S_8$, and the equivalent for the other parameters.  The input values are listed in Table~\ref{tab:priors}, with the $\Delta$ offsets determined from the difference between the mean marginal estimates from the `truth' and `test' mock analyses (see Section~\ref{sec:truthtest} for details).
    For the purpose of this figure, the $0.5\sigma$-width is set to the constraining power of our fiducial analysis. The light grey region corresponds to $\pm 1\sigma$ uncertainty.
    Rows 1-7 review the robustness of the Y1 \hmcode fiducial pipeline, and an alternative \spk-based analysis, to weaker and stronger baryon feedback (Section~\ref{sec:baryon_robustness}) 
    Rows 8-12 assess the impact of systematics mitigation strategies for Y10 (Section~\ref{sec:Y10}).
    Row 13 adopts the alternative TATT intrinsic alignment model for Y1 (Section~\ref{sec:varyIA}). 
    Rows 14-15 vary the Y1 set of cosmological parameters and priors (Appendix~\ref{app:parametersandpriors}). 
    The constraints on $S_8$ are tabulated in Appendix~\ref{app:extrastab}.}
    \label{fig:allresults}
\end{figure*}

We present our forecast $\Lambda$CDM cosmic shear constraints for the parameters $S_8$, $\Omega_{\rm m}$, $A_{\rm IA}$ and $\eta_{\rm IA}$ in Figure~\ref{fig:allresults}.  Our fiducial analysis can be compared to variants of this configuration. 
Throughout this analysis we primarily focus on the $S_8$ parameter, for which our target accuracy is to recover the truth value within $0.5 \sigma$. In Section~\ref{sec:fiducial_results} we define our fiducial pipeline configuration for mitigating the impact of baryon feedback.
In Section~\ref{sec:baryon_robustness} we validate the robustness of of our analysis to `stronger' and `weaker' feedback scenarios, also investigating the alternative \spk method for modelling baryon feedback.
We consider degrading the photometric redshift calibration precision for data at Y10 depth, quantifying the impact this has on our cosmological constraints in Section~\ref{sec:Y10}.
In Section~\ref{sec:varyIA} we consider the impact of our IA model and finally in Section~\ref{sec:w0wa} we consider our $\Lambda$CDM forecasts in the wider context of constraining $w_{\rm 0}w_{\rm a}$CDM with multiple probes.

\begin{table*}
\centering
    
        \begin{tabular}{ll|cr|cr|rcc}
        \toprule
            Baryon Prior & Scale Cut &
            \multicolumn{2}{c|}{`Test' Mock}
            & \multicolumn{2}{c|}{`Truth' Mock}
            & \multicolumn{3}{c}{Bias}  \\
        \midrule 
        & &$S_8$ & $S_8-S_8^{\rm in}$ &$S_8$ & $S_8-S_8^{\rm in}$ &$ \Delta S_8$ & $\sigma_{\rm Test}/\sigma_{\rm Truth}$ & $\sigma/\sigma_{\rm Fid}$\\ [0.2cm] \midrule 
FLAMINGO & $K_{\rm max}=0.5$& $ 0.802^{+ 0.005}_{- 0.006} $ & $ 0.31 \sigma $ & $ 0.799^{+ 0.005}_{- 0.006} $ & $ -0.14 \sigma $ & $ 0.45 \sigma $ & $ 1.04 $ & $ 1.00 $ \\ [+0.1cm] 
        \hline 
FLAMINGO & $K_{\rm max}=1.0$& $ 0.803^{+ 0.005}_{- 0.005} $ & $ 0.52 \sigma $ & $ 0.799^{+ 0.005}_{- 0.006} $ & $ -0.17 \sigma $ & $ 0.69 \sigma $ & $ 0.91 $ & $ 0.92 $ \\ [+0.1cm] 
FLAMINGO & $K_{\rm max}=0.5$& $ 0.802^{+ 0.005}_{- 0.006} $ & $ 0.31 \sigma $ & $ 0.799^{+ 0.005}_{- 0.006} $ & $ -0.14 \sigma $ & $ 0.45 \sigma $ & $ 1.04 $ & $ 1.00 $ \\ [+0.1cm] 
FLAMINGO & $K_{\rm max}=0.1$& $ 0.800^{+ 0.014}_{- 0.012} $ & $ 0.03 \sigma $ & $ 0.800^{+ 0.013}_{- 0.012} $ & $ -0.03 \sigma $ & $ 0.06 \sigma $ & $ 1.03 $ & $ 2.29 $ \\ [+0.1cm] 
        \hline 
DM only & $K_{\rm max}=1.0$& $ 0.793^{+ 0.004}_{- 0.004} $ & $ -1.91 \sigma $ & $ 0.800^{+ 0.004}_{- 0.004} $ & $ -0.10 \sigma $ & $ -1.82 \sigma $ & $ 0.97 $ & $ 0.69 $ \\ [+0.1cm] 
DM only & $K_{\rm max}=0.5$& $ 0.795^{+ 0.005}_{- 0.004} $ & $ -1.01 \sigma $ & $ 0.800^{+ 0.005}_{- 0.005} $ & $ -0.08 \sigma $ & $ -0.93 \sigma $ & $ 1.00 $ & $ 0.82 $ \\ [+0.1cm] 
DM only & $K_{\rm max}=0.1$& $ 0.799^{+ 0.013}_{- 0.012} $ & $ -0.05 \sigma $ & $ 0.800^{+ 0.012}_{- 0.013} $ & $ -0.02 \sigma $ & $ -0.03 \sigma $ & $ 1.00 $ & $ 2.15 $ \\ [+0.1cm] 
        \hline 
Uninformative & $K_{\rm max}=1.0$& $ 0.801^{+ 0.006}_{- 0.007} $ & $ 0.19 \sigma $ & $ 0.799^{+ 0.006}_{- 0.006} $ & $ -0.23 \sigma $ & $ 0.42 \sigma $ & $ 1.06 $ & $ 1.13 $ \\ [+0.1cm] 
Uninformative & $K_{\rm max}=0.5$& $ 0.799^{+ 0.005}_{- 0.008} $ & $ -0.17 \sigma $ & $ 0.797^{+ 0.006}_{- 0.009} $ & $ -0.45 \sigma $ & $ 0.28 \sigma $ & $ 0.93 $ & $ 1.19 $ \\ [+0.1cm] 
Uninformative & $K_{\rm max}=0.1$& $ 0.802^{+ 0.012}_{- 0.013} $ & $ 0.18 \sigma $ & $ 0.802^{+ 0.013}_{- 0.014} $ & $ 0.12 \sigma $ & $ 0.06 \sigma $ & $ 0.95 $ & $ 2.27 $ \\ [+0.1cm] 
        \hline 
DESxROSAT & $K_{\rm max}=1.0$& $ 0.802^{+ 0.005}_{- 0.004} $ & $ 0.52 \sigma $ & $ 0.799^{+ 0.004}_{- 0.004} $ & $ -0.18 \sigma $ & $ 0.69 \sigma $ & $ 1.10 $ & $ 0.77 $ \\ [+0.1cm] 
DESxROSAT & $K_{\rm max}=0.5$& $ 0.802^{+ 0.005}_{- 0.005} $ & $ 0.44 \sigma $ & $ 0.799^{+ 0.005}_{- 0.004} $ & $ -0.15 \sigma $ & $ 0.60 \sigma $ & $ 1.02 $ & $ 0.85 $ \\ [+0.1cm] 
DESxROSAT & $K_{\rm max}=0.1$& $ 0.800^{+ 0.013}_{- 0.012} $ & $ -0.002 \sigma $ & $ 0.799^{+ 0.012}_{- 0.013} $ & $ -0.05 \sigma $ & $ 0.04 \sigma $ & $ 0.96 $ & $ 2.17 $ \\ [+0.1cm] 
        \hline 

        \hline
        \end{tabular}
        
        \caption{LSST-Y1 $\Lambda$CDM cosmic shear forecasts for $S_8$. 
        We consider three sets of scale cuts, $K_{\rm max}=[0.1, 0.5, 1.0]$, and three choices of prior for the \hmcode baryon feedback parameter:  Uninformative, $\Theta_{\rm AGN} \in [6.5, 9.36]$; FLAMINGO-informed, $\Theta_{\rm AGN} \in [7.4, 7.95]$; DESxROSAT, $\Theta_{\rm AGN} \sim {\cal N}(7.694; 0.008)$.  Constraints from a `scale-cut only' analysis, where baryon feedback is not modelled, are also tabulated (DM only). We conduct two analyses for each variant. In the `Truth Mock', the input mock is created by, and analysed with, \hmcode.  The offset from the input cosmology, $S_8 - S_8^{\rm in}$, in this `truth' case, is therefore a measure of projection bias, which varies for each analysis setup, and can be as high as $-0.45\sigma$ when using uninformative priors.  In the `Test Mock', the mock is created by the fiducial FLAMINGO L1\_m9 simulation, but analysed with \hmcode (see Section~\ref{sec:truthtest} for details), creating an offset in the recovered value of $S_8$ which combines both model bias and projection effects. Throughout this study we remove the projection offset by reporting $\Delta S_8 = (S_8^{\rm truth} - S_8^{\rm test})$ as a fraction of the mean marginal $1\sigma$ error estimate. The table also compares the relative constraining power of each analysis as the ratio, $\sigma/\sigma_{\rm Fid}$, to the fiducial $K_{\rm max} = 0.5$ analysis reported in the top row.}
        \label{tab:scalecuts}
\end{table*}

\subsection{Defining the fiducial Y1 analysis}
\label{sec:fiducial_results}
\begin{figure}[h!]
    \centering
    \includegraphics[width=\columnwidth]{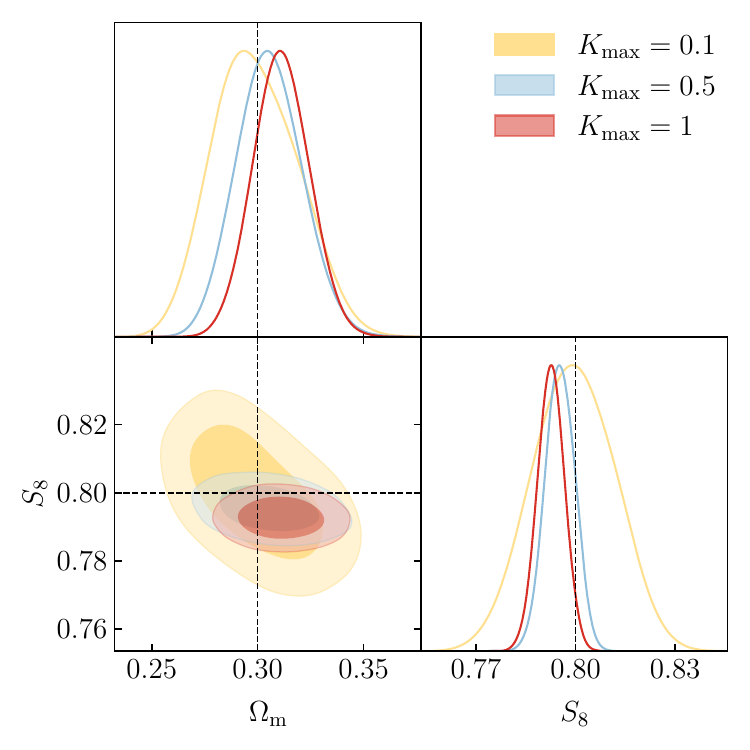}
    \caption{LSST Y1 cosmic shear forecasts on $S_8$ and $\Omega_{\rm m}$ for a `scale cut only' analysis that does not model baryon feedback. 
    We compare three different redshift-dependent scale cuts: $K_{\rm max}=1$ (red),  $K_{\rm max}=0.5$ (blue), and $K_{\rm max}=0.1$ (yellow) (see Appendix~\ref{app:extrastab} for how $K_{\rm max}$ relates to $\ell_{\rm max}$ in different tomographic bins).
    As we remove scales, the constraining power and the bias reduce significantly. 
 A $2\sigma$ bias on $S_8$ for $K_{\rm max}=1$ decreases to a $0.03\sigma$ offset for $K_{\rm max}=0.1$, with the error on $S_8$ tripling and becoming comparable to existing constraints from pre-cursor surveys.   These results are all cases of `test' analyses where the projection `bias' is low at $\sim 0.1\sigma$.    In this figure, and all similar contour plots that follow, the dashed black lines indicate the input cosmology (listed in Table~\ref{tab:priors}), with the marginalised posterior contours showing the 68\% (inner) and 95\% (outer) credible intervals.   
    The constraints on $S_8$ are tabulated in Table~\ref{tab:scalecuts}.}
    \label{fig:no_baryons}
\end{figure}

The \citetalias{DESCSRD/etal:2018} fiducial cosmic shear analysis adopts a scale cut of $\ell<3000$ to mitigate against the bias introduced from the uncertain impact of baryon feedback in a $w_{\rm 0}w_{\rm a}$CDM cosmology. 
In Figure~\ref{fig:no_baryons} we compare the LSST Y1 cosmic shear constraints on $S_8$ and $\Omega_{\rm m}$ for a `scale cut only' flat $\Lambda$CDM analysis, progressively removing small-scale information with redshift-dependent scale cuts that range from $K_{\rm max}=1$ (an average $\ell_{\rm max} \sim 1400$) to $K_{\rm max}=0.1$ (an average $\ell_{\rm max} \sim 140$, see Appendix~\ref{app:extrastab}).
The unacceptable $1.8 \sigma$ bias recovered for the $S_8$ parameter with a $K_{\rm max}=1$ scale cut can be reduced to essentially zero bias by reducing $K_{\rm max}$ to $0.1$. This bias reduction comes at a cost, however, tripling the error on $S_8$ compared to the $K_{\rm max}=1$, recovering a value for $S_8$ that is hardly more constraining than existing results from \citetalias{des/kids:2023}. 
As the cosmological model simulated for our mock data vector adopts the most likely baryon feedback model from the FLAMINGO simulation, these results are not a `worst case scenario'.  We agree with the conclusion of previous studies that a scale cut-only analysis will not improve on Stage III constraints, thought we note that work such as \citet{truttero/etal:2025} and \citet{damico/etal:2026} explores such approaches.

\begin{figure}[h!]
    \centering
    \includegraphics[width=\columnwidth]{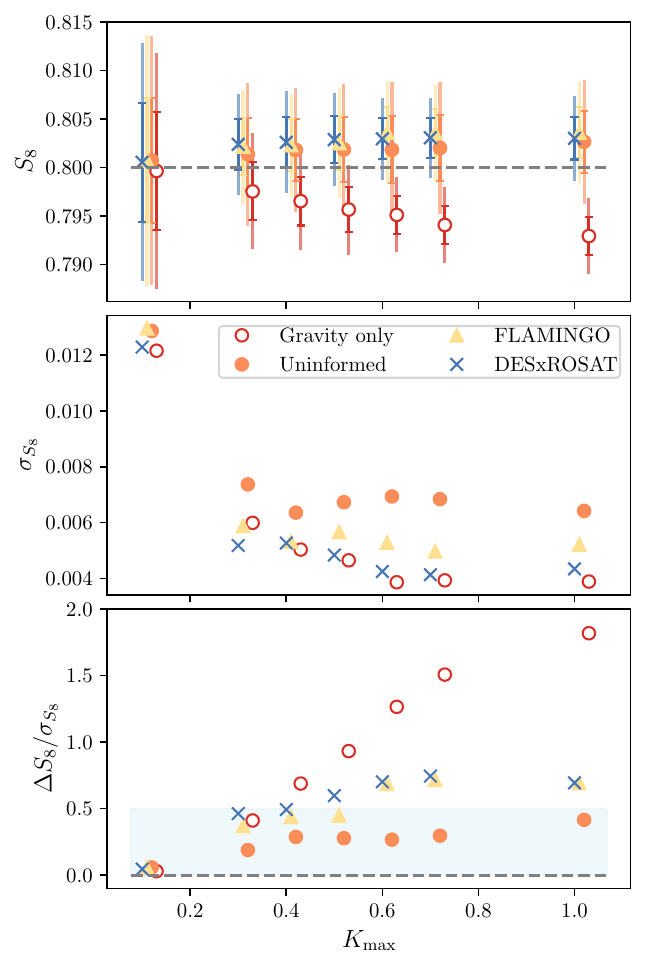}
    \caption{Comparing $S_8$ constraints when varying $K_{\rm max}$ scale-cuts and different baryon feedback priors. 
    Top: Mean marginal `test' analysis constraints on $S_8$, where the inner and outer error bars correspond to $0.5 \sigma$ and $1 \sigma$ respectively, and the dashed line marks the input value. The projection `bias' is reported in Table~\ref{tab:scalecuts}, and can be as high as $0.4\sigma$.
    Middle: 68\% confidence interval error estimate for $S_8$.
    Lower: the absolute value of the offset $\Delta S_8 = |(S_8^{\rm truth} - S_8^{\rm test})|$ as a fraction of the error. We define success as a result where $\Delta S_8< 0.5 \sigma$ (shaded blue).
    We see that a gravity-only analysis that ignores the impact of baryon feedback (red) is only unbiased when excluding all but the smallest $\ell$-scales in the cosmic shear analysis. When using wide priors to marginalise over baryon feedback uncertainty (Uninformed, orange, and FLAMINGO-informed, yellow), little information is gained by increasing $K_{\rm max}$ to high values.  Using external information to constrain the baryon feedback (DESxROSAT, blue) returns precision at the expense of accuracy.}
    \label{fig:scale_cuts}
\end{figure}

For our fiducial analysis we use \hmcode to account for baryon feedback, marginalising over our uncertainty using a single parameter $\Theta_{\rm AGN}$ (see Section~\ref{sec:feedback}). Figure~\ref{fig:scale_cuts} and Table~\ref{tab:scalecuts} compare the recovered constraints on $S_8$ for a range of $K_{\rm max}$ scale cuts and the three different $\Theta_{\rm AGN}$ priors, in addition to the `scale cut only' analysis shown in Figure~\ref{fig:no_baryons}. 

For the uninformative prior with $\Theta_{\rm AGN} \in [6.5, 9.36]$, we meet our $S_8$ accuracy target for all tested scale cuts, but the use of this wide prior significantly impacts our precision. As shown in Figure~\ref{fig:FLAMINGO} and the middle panel of Figure~\ref{fig:scale_cuts}, this wide prior is effectively applying a scale cut by removing all information at high-$\ell$.  The precision on $S_8$ when using an uninformative prior is similar to a scale cut only analysis with $K_{\rm max} \sim 0.2$ $(\ell_{\rm max} \sim 300)$, roughly doubling the error on $S_8$ compared to the `scale cut only' $K_{\rm max}=1.0$ scenario shown in Figure~\ref{fig:no_baryons}.

The use of a highly constraining external prior with $\Theta_{\rm AGN} \in {\cal N}(7.694; 0.008)$ \citep[as motivated by][]{ferreira/etal:2024}, allows us to recover 
most of the precision afforded by a `scale-cut only' analysis. Note that only the width of this Gaussian prior is based on \cite{ferreira/etal:2024} with the centre chosen to match the best fit value of $\Theta_{\rm AGN}$ from an all-scale analysis of our fiducial FLAMINGO mock, when all other parameters are fixed.  Unfortunately this constrained analysis is biased, however, only meeting our threshold criteria that $\Delta S_8 < 0.5\sigma$ for $K_{\rm max}<0.4$ $(\ell_{\rm max} \sim 550)$.  As such it demonstrates the challenges of using tight external constraints on baryon feedback models in cases where the functional form of the model (e.g. \hmcode) is an imperfect reflection of the underlying truth (which in our case is the L1$\_$m9 FLAMINGO simulation). As the physics of baryon feedback is most uncertain at high-$k$, it is plausible that a real-world analysis will also feature a mismatch, a model misspecification, on these scales.    
Our approach of using \hmcode for our forecast is arguably more realistic than one where the analysis model has been trained to perfectly replicate the baryon feedback suppression input to the mock.

For our fiducial analysis set-up we chose a compromise position of adopting a FLAMINGO-based informative prior with $\Theta_{\rm AGN} \in [7.4, 7.95]$, reflecting the fact that we do have some knowledge about the range of plausible feedback scenarios, even if we do not yet know the fine details.  With this prior range we meet our accuracy target for scale cuts $K_{\rm max}<0.5$ $(\ell_{\rm max} \sim 700)$.  This choice affords us a modest 20\% improvement in $S_8$-constraining power, compared to the corresponding analysis with the uninformative prior. The $S_8$ error is, however, a factor of 1.5 larger than the highest precision `scale-cut only' biased constraints.  

Our adopted fiducial scale cut at $K_{\rm max}<0.5$ $(\ell_{\rm max} \sim 700)$ is significantly different from the original \citetalias{DESCSRD/etal:2018} gravity-only forecast which proposed to analyse the cosmic shear power spectra to $\ell<3000$.  Inspecting the middle-panel of Figure~\ref{fig:scale_cuts}, however, we see there is little information gain for $S_8$ beyond $K_{\rm max}=0.6$ $(\ell_{\rm max} \sim 800)$, when considering the highest precision gravity-only analysis.  Taking a pragmatic approach of removing scales that are challenging to model is therefore not as drastic an action as it first appears.

\subsection{Varying the baryon feedback strength and model}
\label{sec:baryon_robustness}
We quantify the sensitivity of our cosmic shear analysis to changes in the input baryon feedback strength by analysing the alternative `weak' and `strong' mock data shown in Figure~\ref{fig:FLAMINGO}. Using our fiducial analysis set-up, a FLAMINGO-based informative prior on $\Theta_{\rm AGN}$ with a scale cut of $K_{\rm max}=0.5$, we recover $S_8$ within our target accuracy of $<0.5 \sigma$, for both extremes in feedback strength. 

\begin{figure}
    \centering
    \includegraphics[width=\columnwidth]{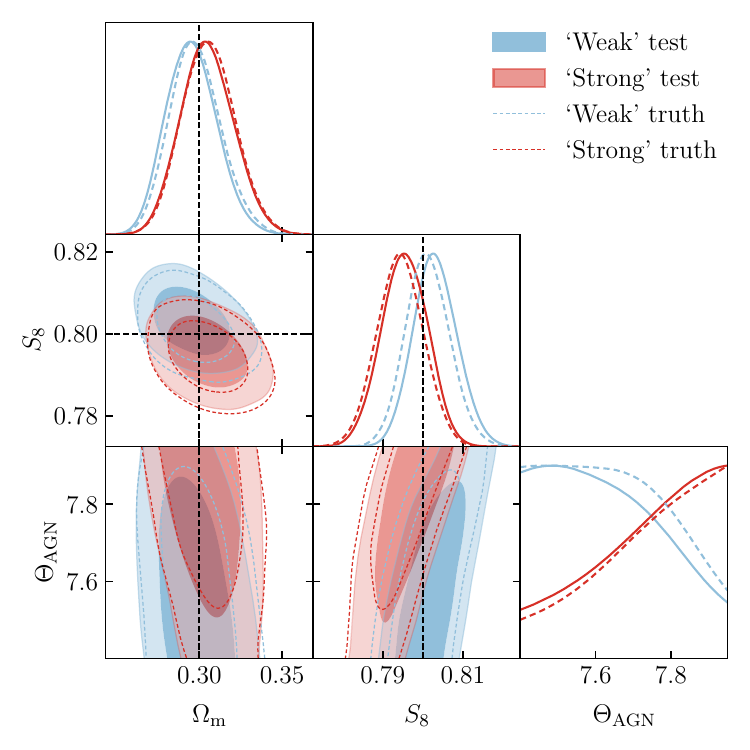}
    \caption{Varying the baryon feedback strength: comparing constraints on $S_8$, $\Omega_{\rm m}$ and the \hmcode baryon feedback parameter $\Theta_{\rm AGN}$ for the weakest (blue) and strongest (pink) feedback scenario from the suite of FLAMINGO simulations.  The `truth' cases quantify the level of `projection bias' relative to the input cosmology (shown dashed).  The `test' cases (where the mock data doesn't match the analysis framework) recover the matching `truth' case within our  target accuracy of $<0.5 \sigma$.    }
    \label{fig:baryon_variants}
\end{figure}

In Figure~\ref{fig:baryon_variants} we compare the `truth' and `test' marginal constraints for the analysis of mock `weak' and `strong' feedback LSST-Y1 cosmic shear. The mean marginal `test' case results (where a FLAMINGO-based mock has been analysed using an \hmcode framework) do not return the input cosmology with $S_8$ offset from the input value by $-0.8\sigma$ and $0.6\sigma$ in the strong and weak scenarios respectively. As seen from the `truth' case results, however, (where strong and weak \hmcode-based mocks have been analysed using an identical \hmcode framework), this offset predominantly arises from projection effects.  The primary source of this `projection bias' is the baryon feedback parameter $\Theta_{\rm AGN}$ which is shown to be correlated with $S_8$.  When informative priors are asymmetrically distributed about the truth, the mean of the one-dimensional marginal posterior distributions for constrained parameters can differ significantly from the input values.  When we account for this projection effect, we can conclude that the accuracy of our fiducial set-up is sufficiently insensitive to changes in the underlying baryon feedback mechanism.

To test the impact of our choice of \hmcode as our feedback model,  Figure~\ref{fig:allresults} compares our fiducial analysis to an analysis that adopts \spk to model the baryon feedback.  Using the same $K_{\rm max}=0.5$ scale cut, this flexible four-parameter baryon feedback model accurately recovers the input $S_8$ values for the fiducial, `weak' and `strong' mock data vectors, with no degradation in constraining power.  The constraints on $S_8$ for these tests are tabulated in Appendix~\ref{app:extrastab}.

\citet{boruah/etal:2024} argue that with a sufficiently accurate and flexible baryon feedback model, the application of any scale cuts is detrimental to the cosmological constraining power.  Based on the highly accurate \spk recovery of the FLAMINGO 3D power-spectra, we therefore conduct an \spk analysis using scales to $\ell_{\rm max}=5000$.  Analysing the full fiducial and `strong' mock data vector with \spk, we are unable to recover $S_8$ at target accuracy, finding a bias of $\sim 0.7\sigma$ (after accounting for projection bias) and a $\sim 15$\% improvement on constraining power compared to our scale cut analysis. The bias can be understood by noting that the \spk emulator is limited to $k<12 h \, {\rm Mpc}^{-1}$, and our cosmic shear $C_\ell$ mocks are projections of the FLAMINGO 3D power spectra to a $k_{\rm max}=40 {\rm Mpc}^{-1}$.  This mismatch could be rectified with a new \spk training sample.  Given the marginal $\sim 15$\% improvement in constraining power, however, we find no strong motivation to extend our analysis beyond \hmcode at this time. Access to tight observational constraints on the $f_{\rm b} (M_{\rm halo})$ relationship \citep[for example, ][]{bigwood/etal:2024}, and hence the four \spk parameters, would, however, clearly motivate the use of \spk over \hmcode in the future \citep[see also][]{vandaalen/etal:2026}.

\subsection{Varying survey depth: analysing Y10}
\label{sec:Y10}
Across ten years of observations, LSST will more than double the effective source density for cosmic shear studies. In this analysis we have adopted the \citetalias{DESCSRD/etal:2018} survey parameters for Y10, assuming a survey with 14,300 square degrees of data down to a 5$\sigma$ limiting magnitude of i < 26.35 (see Table~\ref{tab:mock_ingredients} and Appendix~\ref{app:mock_ingredients}).
This unrivalled depth, however, raises a significant issue for the precise calibration of photometric redshifts without a representative deep spectroscopic sample. 
To account for this challenge we increase our uncertainty on the Y10 redshift accuracy to $\delta_z^i = 0.015$, based on existing analyses that utilise an alternative cross-correlation calibration scheme (as discussed in Section~\ref{sec:datasys}).  Our approach contrasts with \citetalias{DESCSRD/etal:2018}, where the systematic uncertainty requirements were set at a target level  $\delta_z^i = 0.001(1+z_i)$, highlighting the current order of magnitude difference that exists between the redshift calibration accuracy that could be accomplished for Y10 today and the requirements of the future.

\begin{figure}
    \centering
    \includegraphics[width=\columnwidth]{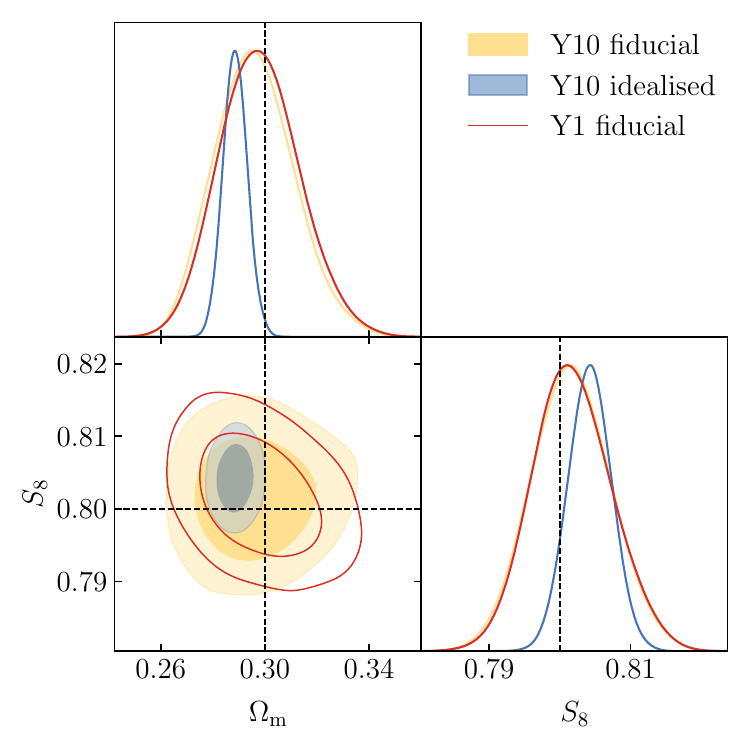}
    \caption{Comparison of fiducial analyses for LSST-Y1 (red) and LSST-Y10 (yellow) mock cosmic shear measurements, showing little improvement in constraining power for Y10 over Y1. Our Y10 baseline analysis increases the uncertainty on the redshift accuracy compared to Y1, with $\delta_z^i=0.015$, to account for the challenges of redshift calibration with an ultra-deep survey. When adopting the idealised approach of \citetalias{DESCSRD/etal:2018}, the error on $S_8$ and $\Omega_{\rm m}$ is roughly halved (blue).}
    \label{fig:Y1_Y10}
\end{figure}

\begin{figure}
    \centering
    \includegraphics[width=\columnwidth]{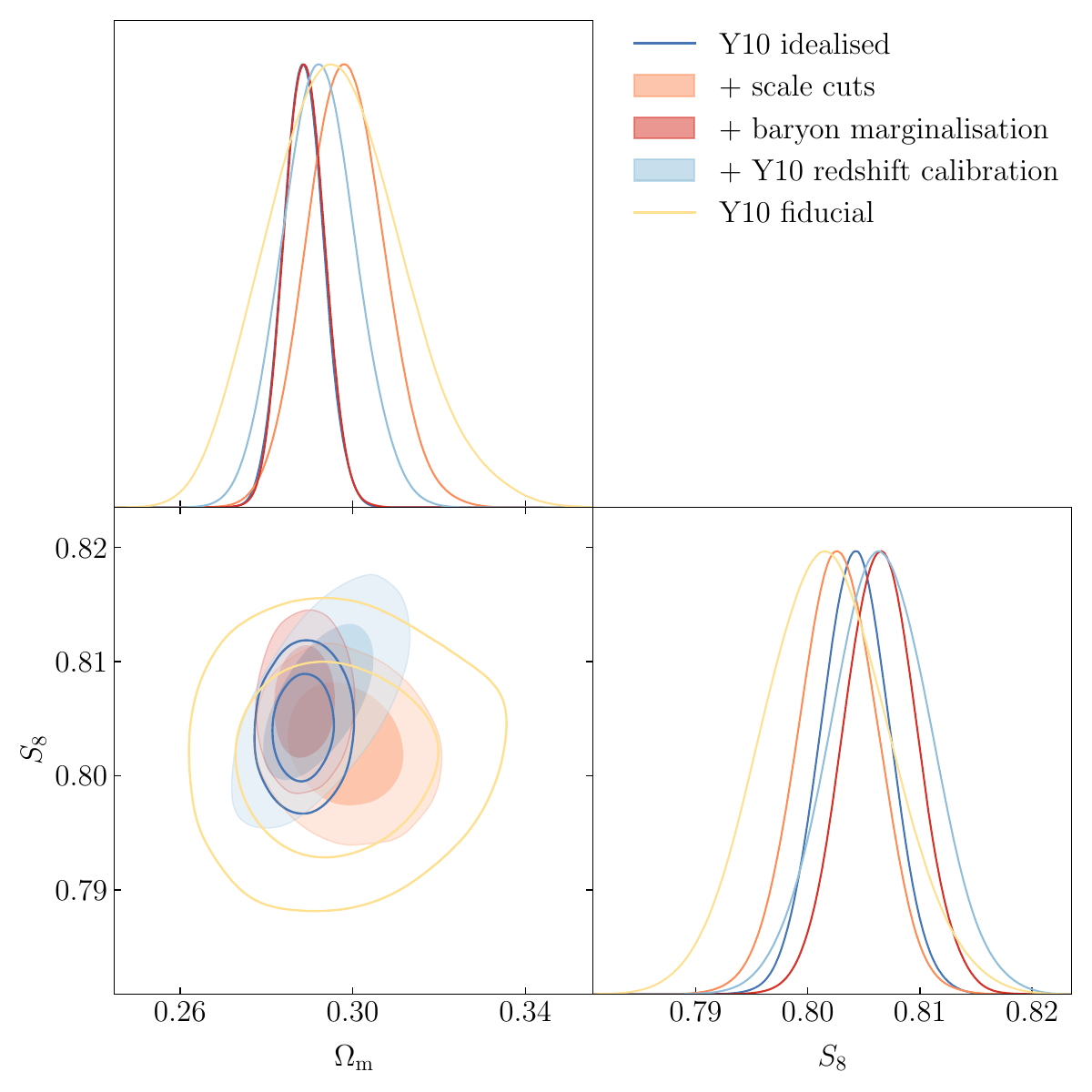}
    \caption{Comparison of the limiting factors in a Y10 cosmic shear analysis. 
    Idealised \citetalias{DESCSRD/etal:2018}-like contours (dark blue, unfilled) are modified by the impact of $K_{\rm max}=0.5$ scale cuts (orange), baryon feedback marginalisation (red) and redshift uncertainty (blue), all of which are combined in our Y10 baseline analysis (yellow unfilled). The constraints on $S_8$ are tabulated in Table~\ref{fig:Y10_limiting_systematics}.}
    \label{fig:Y10_limiting_systematics}
\end{figure}

In Figure~\ref{fig:Y1_Y10} we compare our fiducial analyses of the LSST-Y1 and LSST-Y10 mock cosmic shear measurements, finding little improvement for the Y10 $S_8$ and $\Omega_{\rm m}$ cosmic shear constraint over Y1.  The combination of redshift uncertainty, baryon feedback marginalisation and the $K_{\rm max}=0.5$ scale cuts that we have adopted in our fiducial analyses gives rise to an information plateau that effectively erases the benefits of increasing the survey depth.  To understand the relative impact of these three systematic mitigation strategies for Y10, we run the series of tests summarised in Table~\ref{tab:Y10tests}, with the resulting constraints in the $S_8-\Omega_{\rm m}$ plane shown in Figure~\ref{fig:Y10_limiting_systematics}. 

We start with an `idealised' case, taking the \citetalias{DESCSRD/etal:2018} the science goal requirements on the photometric redshift calibration\footnote{We note that an idealised cosmic shear analysis where the photometric redshift calibration matches the Y1 SOM-based case of $\delta_z^i = 0.002(1+z_i)$ is found to hardly differ from an idealised analysis using the \citetalias{DESCSRD/etal:2018} requirement of $\delta_z^i = 0.001(1+z_i)$.} of $\delta_z^i = 0.001(1+z_i)$; using scales to $\ell_{\rm max}=5000$; fixing the \hmcode baryon feedback amplitude, $\Theta_{\rm AGN}$, at the best-fit value to the fiducial FLAMINGO model.
The error on $S_8$ from this idealised Y10 analysis is almost halved compared to our fiducial case (see Figure~\ref{fig:Y1_Y10}).  Taking this idealised case, we then apply each systematic mitigation strategy in turn. The error on $S_8$ increases by a factor of 1.1 when applying scale cuts, and by a factor of 1.4 when widening the $\delta_z$ prior. Marginalising over baryon feedback has little of impact on the constraining power in the absence of scale cuts.  

We find that none of the Y10 idealised test cases meet our target accuracy to recover the `truth' cosmology within $0.5\sigma$, which is expected based on the corresponding Y1 analysis in Section~\ref{sec:fiducial_results}: with \hmcode we need to combine scale cuts with baryon feedback modelling to recover unbiased results.  It is worth noting that the redshift calibration and baryon marginalisation case find a value for $S_8$ that is biased particularly high.  This results from the wider priors on $\delta_z^i$, or $\Theta_{\rm AGN}$, allowing the nuisance parameters to shift away from their true value in order to absorb the high-$k$ model mismatch between \hmcode and the input FLAMINGO simulation.  \citet{leonard/etal:2024} see similar behaviour when there is a mismatch between the adopted intrinsic alignment model and the input truth.  This demonstrates that any degradation in the precision of the photometric redshift calibration for Y10 over Y1, will further exacerbate the issue of accounting for baryon feedback uncertainty in our cosmic shear analysis.  Improved external observations to constrain these systematic uncertainties will be essential if Y10 is to realise the \citetalias{DESCSRD/etal:2018} forecasts.

\begin{table*}
\centering
    
        \begin{tabular}{l|ccc|ccc}
        \toprule
            Analysis ID & 
            Scale Cuts & $\Theta_{\rm AGN}$ Prior & $\delta_z$ Prior 
        & $S_8$ & $ \Delta S_8$ & $\sigma/\sigma_{\rm Fid}$\\ [0.2cm] \midrule 
Idealised & $\ell_{\rm max} = 5000$ & $7.694$ & $0.001(1 + z)$& $ 0.804^{+ 0.003}_{- 0.003} $ & $ 1.49 \sigma $ & $ 1.00 $ \\ [+0.1cm] 
        \hline 
Baryon Marginalisation & $\ell_{\rm max} = 5000$ & $\in [7.4, 7.95]$ & $0.001(1 + z)$& $ 0.806^{+ 0.003}_{- 0.003} $ & $ 2.21 \sigma $ & $ 0.98 $ \\ [+0.1cm] 
Scale cuts & $K_{\rm max} = 0.5$ & $7.694$ & $0.001(1 + z)$& $ 0.803^{+ 0.004}_{- 0.003} $ & $ 0.87 \sigma $ & $ 1.14 $ \\ [+0.1cm] 
Y10 Redshift Calibration & $\ell_{\rm max} = 5000$ & $7.694$ & $0.015$& $ 0.806^{+ 0.004}_{- 0.004} $ & $ 1.66 \sigma $ & $ 1.36 $ \\ [+0.1cm] 
        \hline 
Y1 fiducial & $K_{\rm max} = 0.5$ & $\in [7.4, 7.95]$ & $0.002(1 + z)$ & $ 0.802^{+ 0.005}_{- 0.006} $ & $ 0.45 \sigma $ & $ 1.80 $ \\ [+0.1cm] 
Y10 fiducial & $K_{\rm max} = 0.5$ & $\in [7.4, 7.95]$ & $0.015$& $ 0.802^{+ 0.006}_{- 0.006} $ & $ 0.42 \sigma $ & $ 1.78 $ \\ [+0.1cm] 

        \hline
        \end{tabular}
        
\caption{To understand the comparable constraining power of Y1 and Y10 for our fiducial analysis configuration, we take an idealised Y10 case, described in the top line, and consider the impact of three systematic mitigation strategies individually.
These are baryon feedback marginalisation with $\Theta_{\rm AGN}$,  scale cuts and redshift calibration uncertainty with a prior on $\delta_z$. 
Each time we change only one aspect of the pipeline set-up, relative to the idealised case. The final lines report the results of our Y1 and Y10 fiducial analyses.
The right side of the table tabulates the $S_8$ `test analysis' constraints, the bias in $S_8$ (including projection effects which are $\sim -0.1 \sigma$ for each case), and the error on $S_8$ relative to the idealised case.
Constraints in the $S_8$-$\Omega_{\rm m}$ plane are shown in Figure~\ref{fig:Y10_limiting_systematics}.  }
\label{tab:Y10tests}
\end{table*}

\subsection{Varying the intrinsic alignment model}
\label{sec:varyIA}
To test the impact of our choice of the two-parameter NLA-z intrinsic galaxy alignment model,  Figure~\ref{fig:allresults} compares our fiducial analysis to an analysis that adopts the five-parameter TATT model \citep{blazek/etal:2019}.
We find no impact on the accuracy or precision of our cosmic shear constraints on $S_8$ and $\Omega_{\rm m}$ from the choice of intrinsic alignment model. 
The full marginalised posteriors are compared in Appendix~\ref{app:extraTATT}.  

These findings differ from a wide range of data analyses that report an error on $S_8$ that grows by up to $\sim 40$\% when using TATT over NLA-z \citepalias[see the discussion in section 3.3 of][and references therein]{des/kids:2023}.
It is, however, in agreement with mock data studies where the intrinsic alignment model has been simulated to match NLA-z \citepalias{des/kids:2023}, as is the case with this study where the three additional TATT parameters have a `truth' value of zero in our mock.

For this lensing-only analysis, we choose to limit our intrinsic alignment study to this single test case as there is already a significant body of work in this area. 
We refer the reader to \citet{samuroff/etal:2024} for a detailed exploration of the significant bias introduced in an LSST Y1-like NLA-z-based analysis when the true intrinsic alignment mechanism is given by an extreme TATT model.  
This result would nominally motivate the use of TATT as the default, but \citet{fischbacher/etal:2023} and \citet{leonard/etal:2024} demonstrate how without tight priors on the photometric redshift calibration uncertainty, weakly constrained TATT parameters can absorb errors in the redshift distribution leading to a bias on the derived cosmological parameters.  With a `$3\times2$pt' joint lensing and galaxy clustering analysis, many of the issues related to intrinsic alignment model bias can, however, be mitigated \citep{leonard/etal:2024,samuroff/etal:2024,navarro/etal:2026}.

\citet{blot/etal:2025} have recently studied the impact of individual nuisance parameter mis-modelling problems for a combined $3\times2$pt and spectroscopic clustering data vector within $\Lambda$CDM for \Euclid. They predominantly find relatively mild posterior effects of around 1$\sigma$, reflecting both the calibrating effect of combining with clustering data, and the already tight prior requirements on end-stage \Euclid results. They find very large ($\approx 5\sigma$) posterior biases, however, when analysing mock data generated without intrinsic alignments using a simple (NLA) model, which should subsume such a configuration. They ascribe this to prior volume effects, but the discrepancy in their located best-fit values suggests this may not be the cause. We do not find corresponding results in our cosmic shear-only analysis.

In this analysis we adopt a set of overlapping tomographic bins, each with a known 7\% outlier population, adopting correlated priors for the photometric redshift calibration uncertainty $\delta_z^i$ in each bin (see Appendix~\ref{app:mock_ingredients} for details). 
In a robustness test of these choices, we find that assuming uncorrelated $\delta_z$ priors in our Y1 fiducial analysis has no impact on our parameter constraints.
Removing the outlier population, and modifying the input model redshift distributions accordingly, does, however, improve the constraints on the intrinsic alignment parameters by $\sim 10$\% for $A_{\rm IA}$ and $\sim 15$\% for $\eta_{\rm IA}$ \citep[see also][]{zhang/etal:2026}. Despite this improved intrinsic alignment model, however, the recovered constraints on $S_8$ remain unchanged.  For the case of a redshift distribution where the outlier population is unknown, we refer the reader to \citep{mill/etal:2025} and section V1.A of \citet{boruah/etal:2024} which both forecast a significant bias in the recovered cosmological parameter constraints, if the outlier population is not known within 5\% accuracy.

\subsection{Joint Constraints in $w_{\rm 0}-w_{\rm a}$CDM}
\label{sec:w0wa}
In our analyses of both the Y1 and Y10 mock cosmic shear observations we have shown that baryon feedback mitigation strategies contribute significantly to the systematic error budget. 
In order to recover accurate cosmological parameters, without external observations to set tight priors on the amplitude and scale-dependence of the baryon feedback mechanism, we are forced to accept a significant reduction in constraining power on the $\Lambda$CDM model.  
This clear result is, however, in stark contrast to the LSST forecasts of \citet{boruah/etal:2024} who find that baryon feedback contributes just 1\% to their total error budget.  In this section we discuss how differences in methodology and success metrics lead to these contrasting conclusions.

Our cosmic shear-only $\Lambda$CDM analysis differs from \citet{boruah/etal:2024} in that they analyse the impact of astrophysical systematics in a `$3\times2$pt' forecast of $w_{\rm 0}w_{\rm a}$CDM. 
Their metric of success is the composite dark energy parameter $w_{\rm p} = w_{\rm 0} + w_{\rm a} [z_{\rm p}/(1+z_{\rm p})]$ at a pivot redshift $z_{\rm p}=0.4$.  
Using a flexible baryon feedback model they analyse the $3\times2$pt LSST data vector, finding no need to introduce any additional scale cuts\footnote{\citet{boruah/etal:2024} use real space statistics, applying a scale cut of 1 arcmin for both components of the two-point shear correlation function $\xi_{\pm}$.} in their analysis as $w_{\rm p}$ is recovered, for a range of input baryon feedback models, within the target accuracy of $< 0.3 \sigma$.    

One concern over the use of the $w_{\rm p}$ metric, is that it is insensitive to biases in other cosmological parameters.  
As the target dark energy constraints for LSST rely on the inclusion of additional and external probes, we consider whether a biased $S_8$ constraint from an inaccurate cosmic shear analysis could introduce an error in the joint dark energy constraints.    

\begin{figure}
    \centering
    \includegraphics[width=\columnwidth]{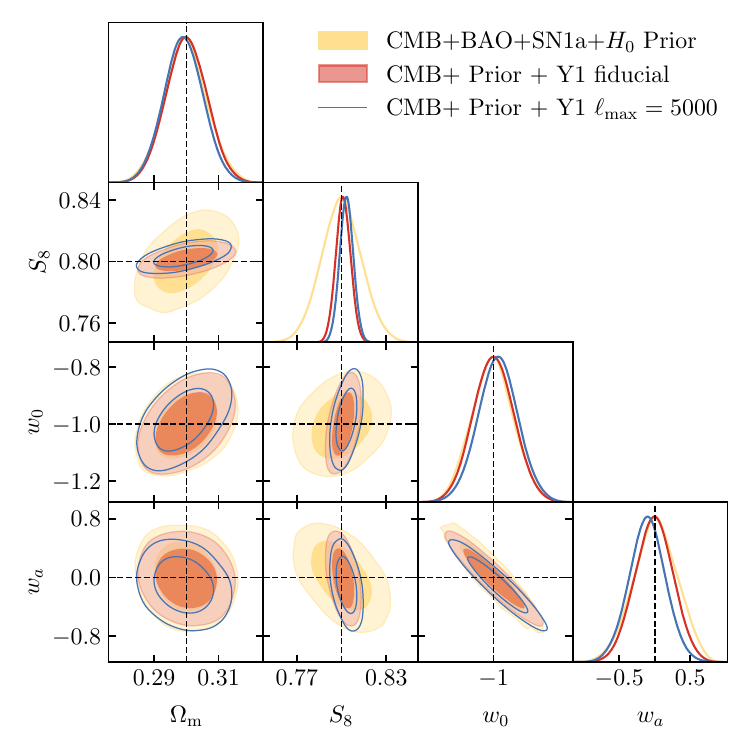}
    \caption{Comparison of constraints on $\Omega_{\rm m}$, $S_8$, $w_{\rm 0}$ and $w_{\rm a}$ for our fiducial Y1 cosmic shear data vector with a `CMB+' prior  (yellow) that imitates combined constraints from the CMB \citep{planck/etal:2020}, baryon acoustic oscillations \citep{alam/etal:2017}, the local distance ladder \citep{riess/etal:2018} and Type Ia supernova \citep{scolnic/etal:2018}.
    By including Y1 cosmic shear data using our fiducial analysis (red), the constraints on $S_8$ are improved by a factor of 2.7. 
    We find only a slight improvement in the constraints on $w_{\rm 0}$ and $w_{\rm a}$ in a joint analysis, even when including small-scale information to an $\ell_{\rm max}=5000$ (blue).  This motivates us to focus on $S_8$ constraints when setting requirements on early LSST cosmic shear analyses. }
    \label{fig:w0wa}
\end{figure}

Figure~\ref{fig:w0wa} presents $w_{\rm 0}w_{\rm a}$CDM forecasts for an analysis of our fiducial LSST-Y1 cosmic shear mock data vector, including a `CMB+' prior that mimics the combined constraints from the `Stage-III' CMB \citep{planck/etal:2020}, baryon acoustic oscillations \citep{alam/etal:2017}, the local distance ladder\footnote{The `Stage-III' prior re-produces the constraints expected from the combination of data sets, assuming there is no tension between the different probes.} \citep{riess/etal:2018} and type Ia supernova \citep{scolnic/etal:2018}.  
We find that the addition of our Y1 fiducial cosmic shear analysis would decrease the error on $S_8$, by a factor of 2.7, over Stage-III alone. 
Very little impact is seen for the constraint on $w_{\rm 0}$, which improves by only a few percent in the joint analysis, but the error on $w_{\rm a}$ decreases by $16\%$.

Following \citet{boruah/etal:2024}, we conduct a joint analysis that includes cosmic shear data to an $\ell_{\rm max}=5000$ and marginalises over our uncertainty. In this `all scale' analysis we find a significant $\sim 0.8\sigma$ bias in $S_8$, as expected from Section~\ref{sec:fiducial_results}. Importantly, however, we find that this offset in $S_8$ does not translate into a large biases in the dark energy parameters.  The bias in the recovered $w_{\rm 0}$ and $w_{\rm a}$ constraints in this case rise to $0.2\sigma$ and $0.4\sigma$ respectively, which is within our level of tolerance. As such we can understand the differing conclusions between this study and \citet{boruah/etal:2024}.  The use of an imperfect baryon feedback mitigation scheme significantly impacts the $S_8$ constraints in a $\Lambda$CDM analysis. For the dark energy parameters in a $3\times2$pt $w_{\rm 0}w_{\rm a}$CDM analysis, however, the impact is relatively small.  We conclude that the requirements on the accuracy of the baryon feedback model can be reduced if the goal of the analysis is to solely constrain the dark energy parameters.

\section{Conclusions}
\label{sec:conclusions}

LSST and \Euclid have been designed to deliver percent-level precision measurements of the $w_{\rm 0}$ and $w_{\rm a}$ parameters that model the evolving equation of state of dark energy. 
To achieve that goal though, both surveys will require additional information from the CMB observations that currently lie in mild tension with a number of existing low-redshift observables. As offsets between complementary probes could be indicative of systematic bias or `new physics', this analysis has focussed on quantifying how significant the bias could be for an LSST cosmic shear analysis in light of the wide range of feasible baryon feedback scenarios recovered from the latest FLAMINGO suite of hydrodynamical simulations \citep{schaye/etal:2023}. 
In order to reduce the systematic bias on the most constrained parameter, $S_8$, to below $0.5\sigma$, we find that our fiducial analysis requires the removal of small-scale information out to a $K_{\rm max}=0.5$ ($\ell_{\rm max} \sim 700$).   
This approach improves accuracy, but at a cost: the $S_8$ error for our unbiased fiducial analysis is almost double that of the $\sim 2\sigma$-biased analysis where baryon feedback is ignored.

In our fiducial analysis we marginalise over the amplitude of the baryon feedback suppression to the non-linear matter power spectrum using the one-parameter \hmcode model \citep{mead/etal:2021}.
This approach is arguably insufficient to model the complexity of the FLAMINGO simulations, but unfortunately our conclusion about the degradation in constraining power that arises from baryon feedback mitigation is unchanged by the adoption of the more sophisticated four-parameter \spk model \citep{salcido/etal:2023}. 
This flexible emulator has been trained on 400 ANTILLES hydrodynamical simulations providing an emulation of all the ANTILLES models, accurate to 2\% out to $k<10 h\,{\rm Mpc}^{-1} $. 
Without tight priors to inform the \spk parameters, however, the marginalisation over the full ANTILLES space effectively removes most of the cosmic shear information at high-$\ell$.  In an `all-scale' \spk cosmic shear analysis, the constraints on $S_8$ are unchanged compared to a scale-cut \spk analysis.  Problematically, this flexible `all-scale' \spk analysis also fails to meet our target accuracy, with $\sim 0.7\sigma$ biases in the recovery of $S_8$, highlighting the challenge of analysing data at the high-$\ell$ scales where there is significant uncertainty over the impact of baryon feedback. 

We find that the use of the DESxROSAT tight prior for the baryon feedback model \citep{ferreira/etal:2024}, reduces the error on $S_8$ to become comparable with the forecasts from the idealised \citetalias{DESCSRD/etal:2018} case where baryon feedback is ignored.  This prior needs to be combined with an accurate scale-dependent model, however, which we currently lack.  Throughout this analysis we have shown how the mild high-$k$ mismatch between our analysis models (\hmcode and \spk) and our mock FLAMINGO input can be absorbed by other parameters in the cosmological analysis, leading to biased constraints on $S_8$.  Unfortunately the solution is not as simple as updating our analysis models to better match FLAMINGO, as recent joint X-ray-kSZ analyses have demonstrated that these state-of-the-art hydrodynamical simulations are likely incomplete, unable to fully reproduce the extended gas profiles seen in observational data \citep{siegel/etal:2026}.  The mismatch between model and mock that we have included in this analysis therefore  provides a reasonable reflection of our current uncertainty of this effect.  We conclude that understanding baryon feedback must become a top priority for both LSST and \Euclid, if cosmic shear is to remain a competitive probe.  

We recognise that our conclusion appears at odds with the LSST forecasts from \citet{boruah/etal:2024} who found that baryon feedback uncertainty contributes minimally to the error budget. This difference results from our two studies addressing two different questions.
In this analysis we have focused on forecasts for $\Lambda$CDM, motivated by the current drive in the cosmological community to understand the tensions that exist between current data sets. 
In contrast \citet{boruah/etal:2024} ask: how accurately will an LSST $3\times2$pt analysis constrain dark energy? 
In our joint analysis forecast of LSST-Y1 cosmic shear with the CMB and other external probes, we also conclude that the baryon feedback bias detected in our $\Lambda$CDM analysis is not detrimental to the joint constraints on $w_{\rm 0}$ and $w_{\rm a}$.  
It is worth highlighting, however, that the $3\times2$pt LSST-Y1 analysis will not produce competitive dark energy constraints compared to the combined constraints from existing surveys \citepalias[see for example figure G2 of][]{DESCSRD/etal:2018}. 
The most interesting science case for early LSST and \Euclid data will therefore be the question of tension, both internal tension between different cosmological probes, and external tension with data from the CMB. 
To ensure confidence in any levels of tension that may be detected, cosmic shear analysis pipelines must become robust to systematic uncertainties within the more constraining $\Lambda$CDM framework that we have studied.
Developing new observational probes \citep[\eg][]{ferreira/etal:2024,mccarthy/etal:2023,bigwood/etal:2024,wayland/etal:2025} and flexible theoretical modelling \citep[\eg][]{debackere/etal:2020, mead/etal:2020, asgari/etal:2023} to constrain baryon feedback will be central to achieving this goal.

We have also reviewed the impact of varying the intrinsic alignment model, finding no significant impact on our constraints. 
This conclusion comes with a caveat, however, as the response of cosmic shear constraints to changes in the analysis model is known to be dependent on the type of intrinsic alignments that have been simulated in the mock data vector \citepalias[see for example][]{des/kids:2023}.
An alternative approach to intrinsic alignment has been demonstrated in \citet{mccullough/etal:2026} and \citet{bigwood/etal:2025b}, where the galaxy sample is limited to only `blue' star-forming galaxies that are less impacted by intrinsic alignments, such that subsequent constraints are more stable to these modelling choices.
In Appendix~\ref{app:parametersandpriors} we demonstrate that the LSST Y1 $S_8$ constraints are insensitive to the choice of the cosmological parameter space and Appendix~\ref{app:truth_params} highlights how adopting a double pass `test-truth' analysis strategy can circumvent the challenges that projection effects introduce when determining a metric of success.

Looking ahead to the conclusion of LSST, we quantify the impact of using a necessarily more complex photometric redshift calibration methodology suited to LSST-Y10. For LSST-Y1, sufficiently deep and representative spectroscopic samples already exist that allow for the use of SOM-based methods to calibrate Y1 photometric redshifts. Moving to the unprecedented wide-field depths of LSST-Y10 however, photometric redshift calibration will have to instead rely on cross-correlation techniques which are currently roughly an order of magnitude less precise than the SOM-based alternative.  
For our fiducial analysis set-up, we find that this significant degradation in photometric redshift calibration accuracy leaves Y10 no more constraining for $S_8$ than Y1. \citet{zhang/etal:2026} highlight that the redshift uncertainty on each tomographic bin has a different impact on the constraints and should therefore be treated more carefully that the standard approach we have taken of defining uniform photometric redshift nuisance parameters in forecasts. Furthermore \citep{dassignies/etal:2025} have demonstrated on a mock Euclid set-up a photometric redshift calibration accuracy which is comparable with the SOM method for Y1 but using the cross-correlation technique.
We leave that extension of our analysis to future work. Nevertheless, this result places a high priority on the development of multi-sample cross-correlation studies which help constrain the changing galaxy bias properties across the full redshift range, enhancing the precision of the resulting photometric redshift calibration \citep{bernstein:2009}.
We find that if we can reach a SOM-based accuracy for the Y10 photometric redshifts, and include tight priors on the baryon feedback, Y10 $S_8$ cosmic shear constraints will be more than five times as constraining as the existing \citet{wright-cosmo/etal:2025} and \citet{DESY6/etal:2026} cosmic shear constraints.

%\newpage
\section*{Author contributions}
NR: Project Lead for both analysis and paper writing.
JZ and CH: Co-wrote paper and assisted with analysis.
IM, JS, MS, MD and JS: Developed the FLAMINGO simulations.
NS: Assisted with SRD setup and covariance matrix, and provided fisher comparisons.
PB: Developed and trained CosmoPower for the analysis.
JP: Developed and conducted the different sampler analysis.
DL: DESC builder, contributed to covariance code TJPCov and as internal reviewer.

\section*{Acknowledgements}

This paper has undergone internal review in the LSST Dark Energy Science Collaboration. % REQUIRED
The internal reviewers were C.D.~Leonard, T.~Eifler, and R.~Mandelbaum. % Optional but recommended

The DESC acknowledges ongoing support from the Institut National de 
Physique Nucl\'eaire et de Physique des Particules in France; the 
Science \& Technology Facilities Council in the United Kingdom; and the
Department of Energy and the LSST Discovery Alliance
in the United States.  DESC uses resources of the IN2P3 
Computing Center (CC-IN2P3--Lyon/Villeurbanne - France) funded by the 
Centre National de la Recherche Scientifique; the National Energy 
Research Scientific Computing Center, a DOE Office of Science User 
Facility supported by the Office of Science of the U.S.\ Department of
Energy under Contract No.\ DE-AC02-05CH11231; STFC DiRAC HPC Facilities, 
funded by UK BEIS National E-infrastructure capital grants; and the UK 
particle physics grid, supported by the GridPP Collaboration.  This 
work was performed in part under DOE Contract DE-AC02-76SF00515.

We acknowledge the following support: the Max Planck Society and the Alexander von Humboldt Foundation in the framework of the Max Planck-Humboldt Research Award endowed by the Federal Ministry of Education and Research (CH);. the UK Science and Technology Facilities Council (STFC) under grant ST/V000594/1 (NR, CH, JZ).

IGM and JS are supported by the Science and Technology Facilities Council (grant number ST/Y002733/1) and have received funding from the European Research Council (ERC) under the European Union’s Horizon 2020 research and innovation programme (grant agreement No 769130).

N\v{S} is supported in part by the OpenUniverse effort, which is funded by NASA under JPL Contract Task 70-711320, `Maximizing Science Exploitation of Simulated Cosmological Survey Data Across Surveys.'

For the purpose of open access, the authors have applied a Creative Commons Attribution (CC BY) licence to any Author Accepted Manuscript version arising from this submission.
\FloatBarrier % niko trying to fix floats.. not sur eif it is working
\section*{Affiliations}
%\scriptsize
$^{1}$ Institute of Astronomy, Royal Observatory Edinburgh, University of Edinburgh, Edinburgh EH9 3HJ, United Kingdom\\
$^{2}$ Ruhr University Bochum, Faculty of Physics and Astronomy, Astronomical Institute, German Centre for Cosmological Lensing, 44780 Bochum, Germany\\
$^{3}$ Waterloo Centre for Astrophysics, University of Waterloo, Waterloo, ON N2L 3G1, Canada\\
$^{4}$ Department of Physics and Astronomy, University of Waterloo, Waterloo, ON N2L 3G1, Canada\\
$^{5}$ School of Mathematics, Statistics and Physics, Newcastle University, Newcastle upon Tyne, NE1 7RU, United Kingdom \\
$^{6}$ Astrophysics Research Institute, Liverpool John Moores University, 146 Brownlow Hill, Liverpool L3 5RF, UK\\
$^{7}$ Center for Cosmology, Department of Physics and Astronomy, University of California—Irvine, Irvine, CA 92697-4575, USA\\
$^{8}$ Department of Physics, Duke University, Science Dr, Durham, NC 27710, USA\\
$^{9}$ Lorentz Institute for Theoretical Physics, Leiden University, PO box 9506, 2300 RA Leiden, the Netherlands\\
$^{10}$ Leiden Observatory, Leiden University, PO Box 9513, 2300 RA Leiden, the Netherlands\\
\bibliographystyle{mnras}
\bibliography{refs} 
\appendix

\section{Cosmic Shear Theory}
\label{app:theory}
The cosmic shear convergence power spectrum, in a spatially flat cosmology, is related to the non-linear matter power spectrum $P_{\rm mm}(k)$, by
\begin{equation}
    C_{\gamma \gamma}^{ij} = p(\ell) \int \mathrm{d} \chi \, \frac{q^{i}(\chi) q^{j} (\chi)}{\chi^2} \, P_{\rm mm}\left( k=\frac{\ell + 1/2}{\chi} , z(\chi) \right) \, ,
\end{equation}
for tomographic bins $i, j$, 3D wavenumber $k$ and 2D multipole $\ell$. 
Here $\chi$ is the comoving distance and the weak lensing window function, $q$ is given by
\begin{equation}
    q^i (\chi) = \frac{3 H_0^2 \Omega_{\rm m}}{2 c^2} \, \frac{\chi}{a(\chi)} \, g^i(\chi) \, ,
\end{equation}
and we adopt the extended first-order  Limber flat-sky approximation \citep{kilbinger/etal:2017} where
\begin{equation}
    p(\ell) = \frac{\ell^4}{(\ell +1/2)^4} \, .
\end{equation}
The lensing kernel for tomographic bin $i$ is given by 
\begin{equation}
    g^i (\chi) = \int_{\chi}^{\chi_{\rm hor}} \mathrm{d} \chi' \, n_{\rm s}^i(\chi')\, \frac{\chi' - \chi}{\chi'} \, ,
\end{equation}
for a source redshift distribution $n_{\rm s}^i(z) =  n_{\rm s}^i(\chi)  \,d\chi/dz$. 

The observed correlation between galaxy shapes includes contributions from both weak lensing by large-scale structure (denoted `$\gamma$'), and the physical intrinsic alignment of neighbouring galaxies (denoted `I'), with the observed cosmic shear power spectrum  given by
\begin{equation}
    C^{ij}(\ell) = C_{\gamma \gamma}^{ij}(\ell) + C_{\rm II}^{ij}(\ell) + C_{\gamma \mathrm{I}}^{ij}(\ell) \, ,
    \label{eqn:c_ell}
\end{equation}
where
\begin{equation}
    C_{\rm ab}^{ij}(\ell) = \int \mathrm{d} \chi \, \frac{W_{\rm a}^{i}(\chi)W_{\rm b}^{j}(\chi)}{\chi^2} \, P_{\rm mm} \left( k=\frac{\ell + 1/2}{\chi}, z(\chi) \right) \, ,
\end{equation}
with the shear kernel $W_\gamma^i (\chi)= q^i (\chi)$. 
In this analysis we adopt the redshift-dependent non-linear linear alignment model (NLA-z), where the intrinsic alignment kernel $W_{\rm I}^i(\chi)$ is given by
\begin{equation}
W_{\rm I}^i(\chi) =  -A_{\rm IA} \left[ \frac{1+z(\chi)}{1+z_0} \right]^{\eta_{\rm IA}} \frac{\overline{C}_1\rho_{\rm crit} \Omega_{\rm m}}{D(z(\chi))} n_{\rm s}^i(\chi) \, . \label{eqn:NLA-z}
\end{equation}
Here $D(z)$ is the linear growth factor, $\overline{C}_1$ is a normalisation constant and $z_0$ is the pivot redshift which we fix at $z_0=0.62$ \citep{bridle/king:2007, joachimi/etal:2011}.   

We calculate the cosmic shear power spectrum, $C(\ell)$, (Equation~\ref{eqn:c_ell}) with {\sc CosmoSIS} using the {\sc project2d.py} module.   
When creating the mock LSST data, the linear matter power spectrum is computed using \camb \citep{lewis/etal:2000,howlett/etal:2012}, the gravity-only non-linear correction is estimated using \hmcode in the `no feedback' mode \citep{mead/etal:2021} and additional suppression from baryon feedback is modelled using data from the \citet{schaye/etal:2023} FLAMINGO hydrodynamical simulation (see Section~\ref{sec:feedback} for details).  

When analysing the mock cosmic shear data we use the {\sc Cosmopower} neural network suite to emulate the non-linear matter power spectrum, $P_{\rm mm}(k,z)$, including baryon feedback \citep{spuriomancini/etal:2022}. 
We train {\sc CosmoPower} within our chosen multi-dimensional parameter space (see Table~\ref{tab:priors}) using 450,000 power spectra calculated with \camb and \hmcode, now in feedback mode.
For our fiducial Y1 cosmic shear analysis in Section~\ref{sec:fiducial_results}, we verify that our parameter constraints using {\sc CosmoPower} are within $0.1\sigma$ from an analysis that uses \camb+\hmcode directly, finding the runtime when using \cosmopower is a factor of four faster. 
We note that for the analyses which extend the parameter space, for example varying neutrino mass in Section~\ref{sec:neutrino} and constraining dark energy parameters in Section~\ref{sec:w0wa}, we revert back to the full \camb and \hmcode-based analysis, rather than re-training the emulator.

\section{Mock LSST Observations}
\label{app:mock_ingredients}
Our mock LSST cosmic shear observations adopt \citetalias{DESCSRD/etal:2018} defined parameters for the Y1 and Y10 survey area, depth, shear and redshift precision, as listed with explicit references for each value in Table~\ref{tab:mock_ingredients}.  We note that a series of survey cadence optimisation studies have recommended a revision to the \citetalias{DESCSRD/etal:2018}-adopted LSST design, increasing the survey area by $\sim 25\%$ with little degradation in depth as a result of improvements in the anticipated system throughput\footnote{The LSST Survey Cadence Optimisation Committee’s Phase 2 Recommendations are published in the internal LSST document \href{https://pstn-055.lsst.io}{PSTN-055}.}.  As our forecasts show that without improvements to our understanding of baryon feedback and redshift calibration, LSST cosmic shear constraints will be systematics dominated, this expected increase in survey area will not alter our conclusions.

We determine the redshift bins by taking the \citetalias{DESCSRD/etal:2018} parameterised redshift distribution for the full Y1 and Y10 samples, split into five equally populated slices and convolved with a Gaussian error distribution of width $\sigma_i =0.05(1+\bar{z_i})$. 
Here $\bar{z_i}$ is the mean redshift of the {\it i}'th bin.  
This approach to developing a redshift bin model does not include `catastrophic outliers', unfortunately resulting in an unrealistically low level of overlap between non-adjacent bins.
For cosmic shear forecasts interested in optimising scale cuts, the use of idealistic clean and narrow photometric redshift bins could be problematic as they allow for scale cuts in the 2D $\ell$-projection to be more closely related to the desired physical scale cut in $k$.

\begin{figure}
    \centering
    \includegraphics[width=1.0\columnwidth]{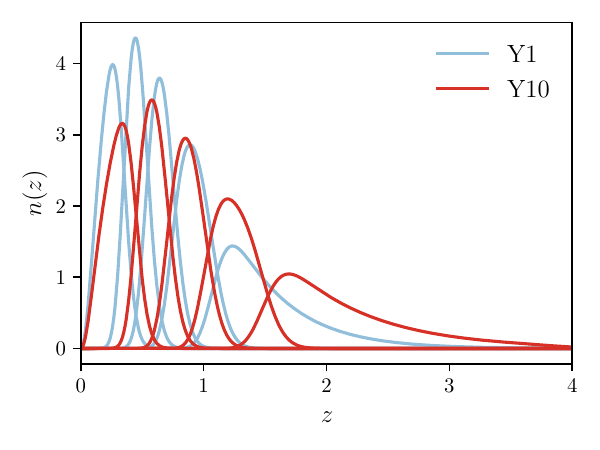}
    \caption{Tomographic redshift distributions used in mock spectrum calculation for our Y1 and Y10. An outlier population is included by spreading 7\% of the sample uniformly in the range $0<z<3$.
    The mean redshift of Y1 is $\langle z \rangle =0.85$ and for Y10 is $\langle z \rangle =1.05$.
    }
    \label{fig:nz}
\end{figure}

LSST catalogue simulations from \citet{graham/etal:2018} derive a photometric redshift outlier population using a colour-space nearest-neighbour technique, finding the majority of outliers lie at either very high (z>2) or very low (z<0.5) redshifts.   
Assuming the shape of this outlier redshift distribution is known, \citet{boruah/etal:2024} find that the outlier fraction must be accurate to within 5-10\% in order to achieve unbiased cosmological parameter constraints.  
In this analysis we do not revisit the question of uncertainty in the outlier fraction. 
Instead, assuming that the shape of the full redshift distribution, including outliers, is known accurately for each bin, we marginalise over our uncertainty on the mean redshift of each tomographic distribution.
We do, however, choose to enhance the realism of the \citetalias{DESCSRD/etal:2018} redshift distribution model by including outliers, adopting a `worse case scenario' for mixing $k$-scales by distributing a 7\% outlier fraction uniformly across the full redshift range from $0<z<3$. 
This model, shown in Figure~\ref{fig:nz}, is motivated by a self-organising map (SOM) analysis of photometric redshifts in the COSMOS field, which results in a 7\% outlier fraction population fairly equally distributed across the full redshift range \citep{stanford/etal:2021}. 
We note that the different outlier population distributions in \citet{graham/etal:2018} and \citet{stanford/etal:2021} could result from their contrasting approaches to photometric redshift estimation.

We calculate the forecast covariance on the LSST cosmic shear power spectrum using {\sc TJPCov}\footnote{The DESC Theory Joint Probe Covariance open source code, {\sc TJPCov}, is available at \href{https://github.com/LSSTDESC/tjpcov}{github.com/LSSTDESC/tjpcov}.}. 
We refer the reader to section 5.2 and appendix E.1 of \citet{joachimi/etal:2021} for a discussion of the various terms that contribute to the covariance. 
In this analysis we include the Gaussian terms which account for shot-noise from the galaxy shear measurements (with $\sigma_e = 0.26$), finite area sampling variance, and the mix between these two error terms.
We also use the {\sc TJPCov} analytic halo model to calculate the super-sample covariance which encodes the impact of Fourier modes larger than the survey window. 
We do not include the non-Gaussian terms, as these are fairly insignificant for cosmic shear. 
Instead the final term we include accounts for the statistical uncertainty in the shear calibration correction, where $\sigma_{\rm m,i} = 0.002(1+\bar{z_i})$ (as discussed in Section~\ref{sec:datasys}), assuming 100\% correlation between the redshift bins \citep[see equation 37 of][]{joachimi/etal:2021}.

\section{Cosmological Parameters and Priors}
\label{app:parametersandpriors}
In every cosmological analysis, there is a choice to be made on the set of cosmological parameters and corresponding priors.
With no standardised method to determine these choices, some surveys choose a parameter space optimised for the observable \citep[see for example the discussion in][]{troester/etal:2020}, with others choosing a $\Lambda$CDM space with different forms for the density parameters (for example $\Omega_{\rm m}$
or $\omega_{\rm c} = [\Omega_{\rm m}-\Omega_{\rm b}]\,h^2$), the power spectrum amplitude parameters ($A_{\rm s}$, $\sigma_8$ or $S_8$), and the dark energy extension. 
Priors can be chosen to be informative based on constraints from complementary analyses.  Alternatively uninformative priors can be selected, although it is challenging to ensure that the complete prior set is truly uninformative\footnote{Consider two cases where a flat prior has been adopted for $\Omega_{\rm m}$ and $\omega_{\rm c}$ respectively. 
Even if the resulting posteriors are insensitive to the breadth of these priors, the different dependencies of the priors on $h$ and $\Omega_{\rm b}$ will influence the posterior's shape. 
The standard statistical approach of determining the uninformative Jeffreys prior does not apply in a multi-dimensional space.  Reference priors \citep{berger/etal:2009,heavens/sellentin:2018} provide a multi-dimensional alternative to Jeffreys, but these have yet to be explored in detail by the cosmological community.}.
In cases of unconstrained parameters with weakly informative priors that are asymmetric around the truth, \cite{chintalapati/etal:2022} and \citetalias{des/kids:2023} demonstrate projection effects within the marginal posterior distributions of constrained parameters, quantifying tension between 1D-marginal statistics and the underlying true cosmology. 
The definition of the cosmological set of parameters and priors, therefore matters, when comparing constraints from different surveys.

\begin{figure*}
    \centering
    \includegraphics[width=\textwidth]{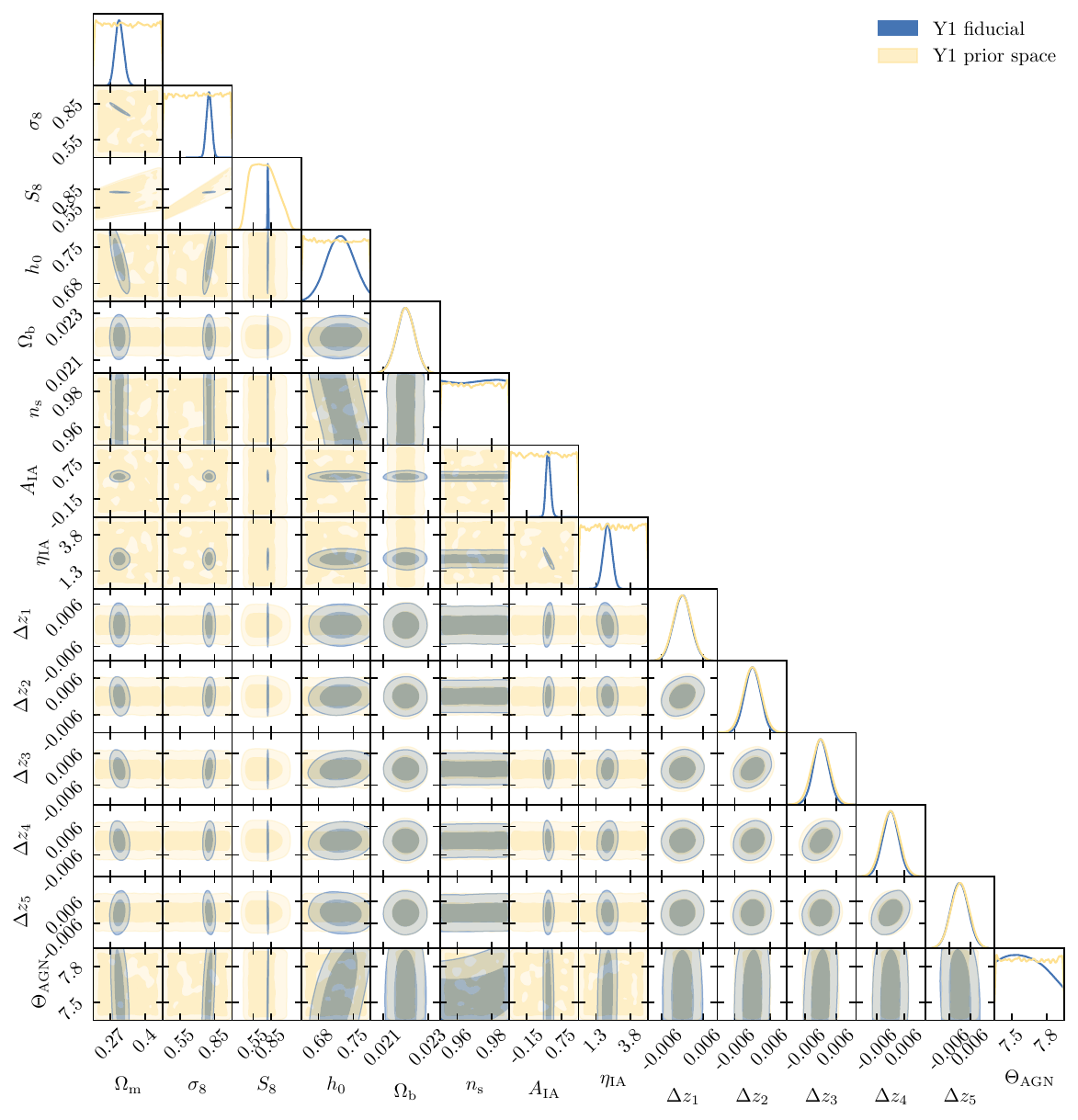}
    \caption{Comparison of samples from the prior (green) and posterior constraints (blue), under our fiducial analysis.
    Only $\Omega_{\rm m}$, $\sigma_8$, $S_8$, $A_{\rm IA}$ and $\eta_{\rm IA}$ are likelihood-dominated; the others are prior-dominated.  
    The choice of priors on the other parameters is important, however; as shown in Figure \ref{fig:DESY3_priors}: some choices can lead to apparent constraints not driven by data.}
    \label{fig:prior_space}
\end{figure*}

In Table~\ref{tab:priors} we list our parameter set and priors, comparing our choices to \citetalias{DESCSRD/etal:2018} which are more informative in all but the notable case of the dark energy parameters. 
We adopt uninformative priors for the cosmic shear-constrained parameters: $\Omega_{\rm m}, \sigma_8, A_{\rm IA}, \eta_{\rm IA}$. 
All other parameters have informative priors, as shown in Figure~\ref{fig:prior_space} where we compare the multi-dimensional prior volume with the posterior from the fiducial analysis presented in Section~\ref{sec:fiducial_results}.  

To set the limits of our uninformative priors we adopt tophat priors, using literature-defined $\pm 7\sigma$ regions for the breadth.  For $\Omega_{\rm m}$, we take constraints from the {\it Pantheon+} analysis of supernova type Ia \citep{brout/etal:2022}, defining a tophat prior with $\pm 7\sigma$ {\it Pantheon}-informed edges of $\Omega_{\rm m} \in \bb{0.2,0.46}$. 
For $\sigma_{8}$, we adopt a $\pm 7\sigma$ region defined by the $3\times2$pt analysis of DES Y3 \citep{desy3/etal:2022} with $\sigma_{8} \in \bb{0.39,1.01}$. 
Note that a $\pm 7\sigma$ region defined by \citet{planck/etal:2020}, in either parameter, would be informative in our analysis.

Setting informative priors requires some subjectivity. 
For $\omega_{\rm b}$ the choice is relatively clear, with two independent observables reporting similar results: $\omega_{\rm b}^{\rm CMB} = 0.02237 \pm 0.00015$ \citep{planck/etal:2020} and $\omega_{\rm b}^{\rm BBN} = 0.02233 \pm 0.00036$ \citep{mossa/etal:2020}. 
We take the more conservative of the two using a Gaussian prior $\omega_{\rm b} \in {\cal N}(0.02233,0.00036)$, sampling in $\omega_{\rm b}$ rather than $\Omega_{\rm b}$ to ensure that this externally constrained prior remains separate from our uncertainty over the value of $h$ where there are a number of disparate constraints. 
Constraints from the CMB favour low values of $H_0$ with $H_0^{\rm CMB} =67.36 \pm 0.54$ \citep{planck/etal:2020}.  Direct measurements from Cepheid observations favour higher values with $H_0^{\rm SH0ES}=73.15 \pm 0.97$ \citep{riess/etal:2022}.
We choose to use a tophat prior with the lower edge defined as $-5\sigma$ from the {\it Planck} best-fit, and the upper edge defined as $+5\sigma$ from the SH0ES constraint, $h:\bb{0.65,0.78}$. 
For $n_{\rm s}$ we motivate the width of the prior set using the $\pm 5\sigma$ region of the $n_{\rm s}$ constraint from \citet{planck/etal:2020}.
We also include additional prior knowledge here by centering our prior around the theoretical expectation for $n_{\rm s}$ such that $n_{\rm s}: \bb{0.95, 0.99}$.
 
\subsection{Sensitivity to the parameter set: $\sigma_8$ or $A_{\rm s}$}
Cosmic shear constraints from existing surveys have been shown to be sensitive to the choice of the power spectrum amplitude parameter within the sampling set \citep[see for example][]{chang/etal:2019}. 
Traditionally cosmic shear surveys sampled using flat priors for either $A_{\rm s}$ or $\ln A_{\rm s}$, but this approach is no-longer favoured\footnote{We note that \citet{sugiyama/etal:2020} derive a correction weight to convert a chain that has been efficiently sampled using uninformative flat priors in $(\Omega_{\rm m},A_{\rm s})$, in order to achieve the desired uninformative priors in $(\Omega_{\rm m},\sigma_{8})$ in the cosmic shear analysis of HSC \citep{dalal/etal:2023}. 
We experimented using this approach but chose not to adopt it. 
As the proposed weighting scheme is only correct to first-order, we found that it lead to a significant slope in the effective $\Omega_{\rm m}$-prior without significantly resolving the issue of the artificial truncation of the $\Omega_{\rm m}-\sigma_8$ degeneracy highlighted by \citet{joachimi/etal:2021}.}, given the non-trivial way a nominally uninformative prior on $A_{\rm s}$ or $\ln A_{\rm s}$, informs both $\Omega_{\rm m}$ and $\sigma_8$ \citep{joachimi/etal:2021}. 
The alternative approach of sampling in $S_8 = \sigma_8 (\Omega_{\rm m}/0.3)^\alpha$, where $\alpha = 0.5$, ensures uninformative priors on this composite parameter, but given the uncertainty on the optimal $\alpha$-value for higher-redshift Stage-IV surveys \citep{wright-cosmo/etal:2025}, we have chosen to sample over $\sigma_8$ for our fiducial analysis.   Interestingly, we found that this choice was rather a moot point: in a robustness test of our fiducial analysis, where we sampled over $A_{\rm s}$, we found less than $0.1 \sigma$ differences between our constrained parameters with the associated uncertainty differing by less than a few percent. Within the statistical power on LSST, we can therefore conclude that the cosmic shear constraints are insensitive to the parameter chosen to model variations in the power spectrum amplitude.

\subsection{Sensitivity to the informative prior width: $h_0$ and $n_{\rm s}$}
\begin{figure}
    \centering
    \includegraphics[width=\columnwidth]{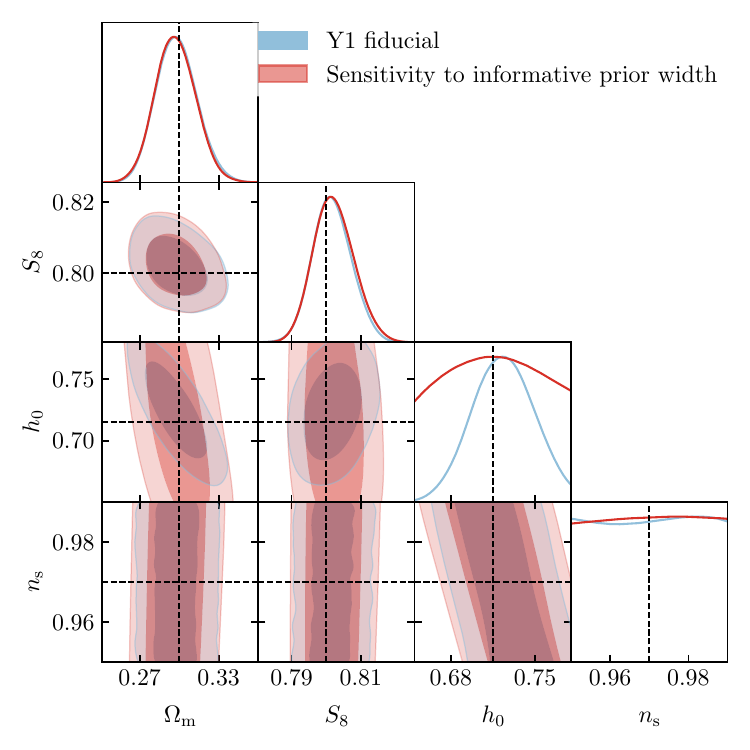}
   \caption{Constraints comparing the Y1 fiducial analysis (blue) to an analysis with less informative priors on the unconstrained parameters $\Omega_b$, $h_0$, and $n_{\rm s}$ (red).
   We find a 4\% increase in the uncertainty on $S_8$ under the broader priors. 
   The $h_0-n_s$ constraints highlight the degeneracy between these two parameters in cosmic shear, such that a strongly informative prior on $n_s$ will result in $h_0$ apparently being constrained, and vice versa. 
   }
    \label{fig:DESY3_priors}
\end{figure}

In Figure~\ref{fig:DESY3_priors} we compare our fiducial analysis from Section~\ref{sec:fiducial_results} with an alternative analysis that adopts the cosmological parameter priors from \citet{amon/etal:2022}. 
Their priors on the unconstrained set of parameters are significantly wider than our fiducial set-up in Table~\ref{tab:priors}, with $h_0 \in [0.55, 0.91]$, $n_{\rm s} \in [0.87, 1.07]$, and $\Omega_{\rm b} \in [0.03, 0.07]$. 
Despite increasing the prior width, these priors nevertheless remain informative, as these particular parameters cannot be constrained by cosmic shear alone.
We find the $S_8$ constraining power reduced by $\sim 4$\% when adopting this weaker set of priors, resulting from weak degeneracies in $h_0 -\Omega_{\rm m}$ and $n_{\rm s}-S_8$. 
In principle the informative priors on these parameters could be extended further, continuing to expand the allowed constrained region in the projected $S_8-\Omega_{\rm m}$ plane. 
But as these regions in the multi-dimensional parameter space can be easily ruled out from other data sets, our preference is to use the tighter informed priors listed in Table~\ref{tab:priors}. 

It is worth noting the strong degeneracy between $h_0$ and $n_{\rm s}$ in Figure~\ref{fig:DESY3_priors}. 
This feature inadvertently leads to an artificial constraint on $n_{\rm s}$ being reported in the LSST-Y1 cosmic shear forecast of \citet{prat/etal:2023}. 
In that case, a strongly informative prior placed on $h_0$ introduced what appeared to be a data constraint on the weakly informed $n_{\rm s}$ parameter. 
With a different set of priors, the converse could also have occurred making it appear that $h_0$ was data-constrained.
We caution that degeneracies between all parameters should be carefully reviewed within the prior volume (see Figure~\ref{fig:prior_space}) before concluding that constraints are data-driven. 

\subsection{Varying the sum of neutrino masses }
\label{sec:neutrino}
Cosmic shear analyses are fairly insensitive to variations in the sum of neutrino masses, especially when applying scale cuts and/or marginalising over a range of baryon feedback scenarios \citep{spuriomancini/bose:2023}.
In our fiducial analysis we therefore adopt the theoretically preferred value for $\Sigma m_\nu=0.006$eV, testing the robustness of this decision with a test analysis where we adopt a top hat prior of $\Sigma m_\nu \in [0.05, 0.6]$eV following \citet{amon/etal:2022}. 
As expected, $\Sigma m_\nu$ is unconstrained by our analysis, and the inclusion of this additional parameter has no significant impact on the precision of $S_8$ or $\Omega_{\rm m}$ measurements. We draw the same conclusions as \citetalias{des/kids:2023}, however, that the introduction of a free $\Sigma m_\nu$ parameter significantly shifts the position of the mean and maximum of the marginal $S_8$ distribution as a result of projection effects \citepalias[see the discussion in section 3.2 of][]{des/kids:2023}.  Our `truth' analysis quantifies the projection `bias' in this case, finding that it shifts the mean marginal to $\sim 0.8 \sigma$ below the input cosmological value.

\section{Sampler Choice}
\label{app:sampler}
In this section, we compare different sampling algorithms available through {\sc CosmoSIS}\footnote{{\sc CosmoSIS}: \href{https://github.com/cosmosis-developers/cosmosis}{github.com/cosmosis-developers/cosmosis}} for characterising our fiducial LSST Y1 model posteriors. 
There are many different ways to sample from a probability distribution; most cosmological analyses have used either \textit{Markov Chain Monte Carlo} (MCMC) methods, especially the Metropolis-Hastings \citep{metropolis/1953, hastings/1970} and \textit{nested sampling} \citep{skilling/2006} algorithms.
Like many recent cosmic shear studies, we focus here on nested sampling methods since they generally require less tuning for specific problems, and also provide the Bayesian Evidence quantity that can be used for model comparisons. 

Nested sampling (NS) is designed for exploring complex, multi-modal likelihoods and for model comparison.
Generically, it integrates from the bounds of a prior distribution in towards the peaks(s) of the posterior, using an ensemble of \textit{live} points in the parameter space. 
By successively replacing the lowest likelihood point and replacing it with a new point, it attempts to evenly sample the subspace of increasingly higher likelihood.

There are many different NS algorithms and software packages; see \citet{buchner/2023} for a recent review.
Here we compare the behaviour of three NS algorithms for our use case: \textsc{MultiNest}\footnote{{\sc Multinest}:\href{https://github.com/farhanferoz/MultiNest}{github.com/farhanferoz/MultiNest}}, \textsc{PolyChord}\footnote{{\sc Polychord}:\href{https://github.com/PolyChord/PolyChordLite}{github.com/PolyChord/PolyChordLite}}, and \textsc{Nautilus}\footnote{{\sc Nautilus}:\href{https://github.com/johannesulf/nautilus}{github.com/johannesulf/nautilus}}. In each case we use the CosmoSIS default parameters for these samplers\footnote{\url{https://cosmosis.readthedocs.io/en/latest/usage/samplers.html}} except as specified below.

\textbf{\textsc{MultiNest}} is an NS algorithm originally designed for multi-modal distributions and curving degeneracies in high dimensions {\citep{Mult_Feroz09}.
It builds ellipsoids around the set of live points and samples within them.
However, as the number of dimensions of a problem increases its number of likelihood evaluations increases exponentially \citep{Poly_Handley15}. Here we use \textsc{Multinest} with 675 live points and efficiency 1.0.

\textbf{\textsc{PolyChord}} uses \textit{slice sampling} \citep{neal/2003}, generating points within 1D slices of constant posterior in the multi-dimensional parameter space \citep{Poly_Handley15}. 
It can take a long time to complete but generally scales well with dimensionality. Here we use \textsc{Polychord} with 250 live points and 30 repeats.

\textbf{\textsc{Nautilus}} uses deep learning to build a model of where to propose new live points, based on all previously sampled points \citep{Naut_Lange23}. 
This algorithm runs in parallel, which is beneficial for large data sets. We use \textsc{Nautilus} with 2000 live points, 512 batches, and 16 networks.

We ran each algorithm on our fiducial model on two 128-core nodes on the National Energy Research Scientific Computing Center (NERSC) machine \textit{Perlmutter}, with 64 processes and four threads per process. We ran our fiducial analysis configuration (which uses \textsc{CosmoPower}) as described in Sec.~\ref{sec:fiducial_results}.
In this configuration,  \textsc{MultiNest} completed in 0.83 hours,  {\sc PolyChord} in 6.2 hours, and \textsc{Nautilus} in 0.86 hours. We use the default stopping criteria for each sampler, which are all slightly different, so potentially any could have attained similar results faster with careful tuning, but the large difference between {\sc PolyChord} and the other two would almost certainly remain.
Contours in the primary $\Omega_m - \sigma_8$ space for the three models agreed very well; 68\% and 95\% chain quantiles of selected
marginalised parameter values for the three methods are shown in Figure~\ref{fig:samplers95}.

All samplers can have different levels of success when applied to new posteriors, requiring tests to be performed on sampling methods when beginning any large analysis campaigns. 
Late Stage III $3\times2$pt posteriors, for example, have proven particularly problematic for recent sampling methods \citep{sanchez-cid/etal:2026}. 
For our analysis, all three samplers recover the same constraints.
We choose not to use the slightly faster \textsc{MultiNest} sampler, as this has been shown in other studies to underestimate the size of the credible intervals in less constraining cosmic shear studies (see appendix D of \citetalias{des/kids:2023} and \citealt{lemos/etal:2023}).  
Given its speed advantage over \textsc{PolyChord}, and the tests against other samplers reported in \citet{Naut_Lange23}, we adopt \textsc{Nautilus} for this project.

\begin{figure}[h]
    \centering
    \includegraphics[width=0.49\textwidth]{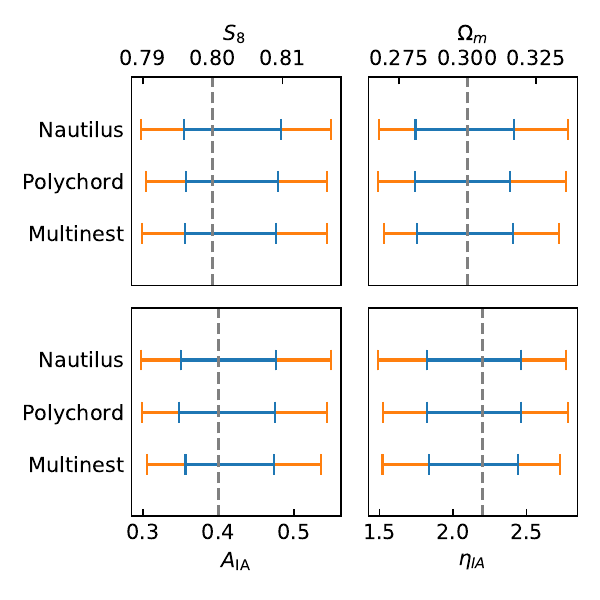}
    \caption{68\% and 95\% quantiles of selected parameters under our fiducial model compared using the \textsc{PolyChord}, \textsc{MultiNest}, and \textsc{ Nautilus} sampling algorithms.
    In our scenario the three methods perform equally well, producing near-identical contours for all parameters of interest.
    We adopt \textsc{Nautilus} for the other chains in this work because of its speed and robustness.}
    \label{fig:samplers95}
\end{figure}

\section{Mitigating projection bias}
\label{app:truth_params}

We optimise our analysis choices based on the accurate recovery of $S_8$, as determined by the difference between the projected mean marginal constraint on $S_8$ from a `truth' and `test' analysis.  In Table~\ref{tab:truthtest} we list the \hmcode and \spk baryon feedback parameters used to create `truth' mocks for the three `test' FLAMINGO models.  These parameters were found to provide the best-fits to the FLAMINGO-created LSST-Y1 cosmic shear power spectra, when analysing all scales with all other parameters fixed to the input values listed in Table~\ref{tab:priors}.

\begin{table}
\centering
        \begin{tabular}{lc|rrr}
        \toprule
	\multicolumn{2}{r|}{Feedback strength}& Weak & Fiducial & Strong \\
	\multicolumn{2}{r|}{FLAMINGO ID}& fgas+2$\sigma$ & L1$\_$m9 &Jet\_fgas-4$\sigma$ \\
	\midrule 
	\hmcode: & $\Theta_{\rm AGN}$ & 7.562 &  7.694 & 7.931\\
	\midrule
	\spk: & $\epsilon$ & 0.350 & 0.297 & 0.251\\
	& $\alpha$ & 0.340 & 0.340 & -0.120\\
	& $\beta$ & 0.770 & 0.770 & 0.740 \\
	& $\gamma$ & 1.190 & 1.13 & 1.020\\
	& $\log(m_{\rm pivot})$ & 13 & 13 & 13 \\
\bottomrule
        \end{tabular}
        \caption{The baryon feedback parameters used to create `truth' mocks for the three `test' FLAMINGO models.}
        \label{tab:truthtest}
\end{table}

In the development of this work we also explored the alternative approach of using the MAP (maximum a posteriori) to define success. In the case of a `truth' analysis, a noise-free MAP estimate will return the input cosmological parameters, arguably making this summary statistic optimal for the assessment of parameter bias without any complications arising from projection bias. To obtain the MAP, we used the {\sc maxlike} sampler within {\sc CosmoSIS} which performs an optimisation process to find the peak of the likelihood.  With the results of a {\sc Nautilus} MCMC analysis to define the starting position, we ran the {\sc maxlike} sampler 20 times. The set of parameters corresponding to the maximum posterior value from those 20 estimates can then be defined as the MAP.  

We analysed several sets of these 20 noisy MAP $S_8$ estimates and noted significant $\sim 0.5\sigma$ differences, even though there was negligible change in the posterior value.  We concluded that as cosmic shear only constrains $S_8$ and $\Omega_{\rm m}$, the {\sc maxlike} sampler's exploration of the wide regions of unconstrained parameter space leads to real variations between the estimated $S_8$ location of the MAP.  As our metric of success is tied to the accurate recovery of the single most-constrained parameter, and not the accurate recovery of all parameters, we concluded that the MAP was therefore not the most appropriate statistic to use for our study.

\section{Additional Data Tables and Figures}
\label{app:extras}

\subsection{Modelling intrinsic alignments with TATT}
\label{app:extraTATT}
Figure~\ref{fig:IA_TATT} shows constraints when analysing our fiducial mock, which includes NLA-z intrinsic alignments, with the extended TATT IA model. 
TATT is a super-set of NLA-z, which adds a tidal torquing alignment mechanism with new amplitude, $A_{\rm IA,2}$, redshift scaling $\eta_{\rm IA,2}$ and a linear galaxy bias amplitude, ${\rm bias_{ta}}$ parameters for the new component \citep[see equations 21-26 in][for implementation]{secco/etal:2022}. 
Our TATT parameter priors follow \citet{desy3/etal:2022}, which are uninformative for LSST-Y1, except in the case\footnote{For cases where the true value of either intrinsic alignment amplitude $A_{\rm IA,i}=0$, it becomes challenging to set a lower limit on the corresponding redshift scaling parameter $\eta_{\rm IA,i}$.  For a low value of $A_{\rm IA,i}$, a large negative value for $\eta_{\rm IA,i}$ can remain acceptable as it only serves to alter the strength of the intrinsic alignment model at very low redshift where the signal-to-noise of the data is insufficient to rule out the model.  For this reason the lower bound of the posterior for the TATT parameter $\eta_{\rm IA,2}$ will not close in an NLA-z mock analysis, even with an extended prior.} of $\eta_{\rm IA,2}$.  
In Figure~\ref{fig:IA_TATT} we can see that the use of the more flexible model has no effect on $S_8$ constraints; the same is also true of the other primary cosmological parameters.
This does not indicate that TATT should be used in place of NLA in all analyses; while it is a more physically motivated model that can capture more general galaxy alignment behaviour, \citet{samuroff/etal:2024} have shown that this can lead to parameter biases in the presence of photometric redshift uncertainty.

\begin{figure}
\centering
\includegraphics[width=\columnwidth]{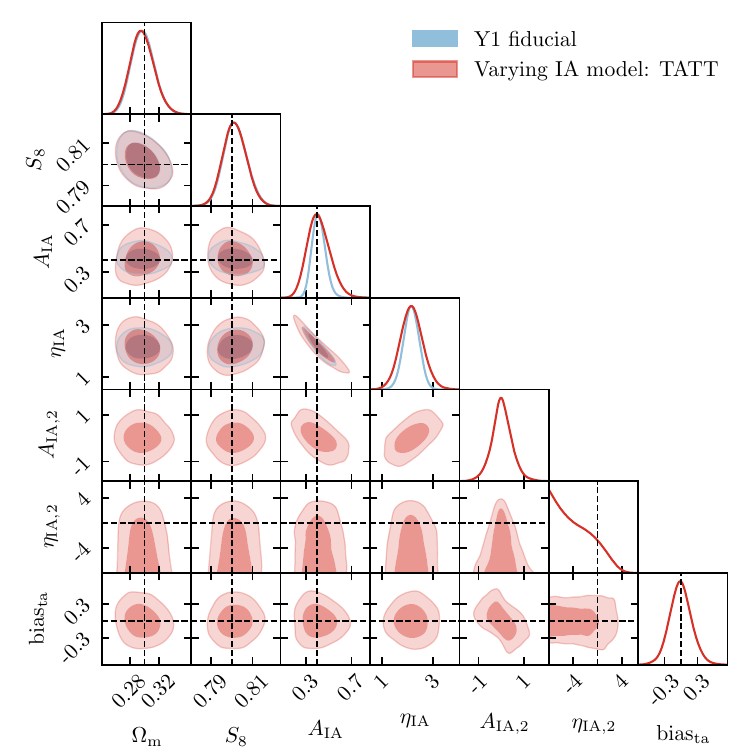}
\caption{Constraints on our primary parameters under our fiducial NLA-z IA analysis (blue) versus an extended TATT analysis, both of an NLA-z mock. 
    Switching to the broader TATT model has no effect on the $S_8$ precision or accuracy.
    The more flexible TATT model can, however, absorb any errors in photometric redshift modelling, leading to cosmological biases.
    }
\label{fig:IA_TATT}
\end{figure}

\subsection{Data tables}
\label{app:extrastab}
We express our scale cut choices as cuts in wave-number $k$, which we then translate to per-bin angular scale cuts using the corresponding distance in our truth model to the median redshift of each tomographic bin (see Equation~\ref{eqn:Kmax-lmax}). 
Table \ref{tab:scalecuts} lists the $\ell$ cuts for each $K_{\rm max}$ cut that we consider.  Tables \ref{tab:baryon_detail} collects measurements of precision and accuracy metrics for different baryon feedback simulations in the Y1 mock and two different methods for modelling, \hmcode and \spk.

\begin{table*}
\centering
        \begin{tabular}{ll|cccccc}
        \toprule
            $z-$bin & $\langle z \rangle$ 
            & \multicolumn{6}{c}{$\ell_{\rm max}$}  \\
        \midrule 
        &  &$K_{\rm max}=0.1$ & $K_{\rm max}=0.2$ & $K_{\rm max}=0.4$ & $K_{\rm max}=0.5$ & $K_{\rm max}=0.6$ &$K_{\rm max}=1$\\ [0.1cm] \midrule 
1 & 0.26 & 83 & 165 & 331 & 413 & 496 & 827 \\ [+0.1cm]
2 & 0.46 & 120 & 240 & 480 & 600 & 720 & 1200 \\ [+0.1cm]
3 & 0.66 & 143 & 287 & 574 & 717 & 861  &1434 \\ [+0.1cm]
4 & 0.92 & 161 & 323 & 646 & 807 & 969 & 1615 \\ [+0.1cm]
5 & 1.42 & 174 & 348 & 696 & 870 & 1044 & 1741 \\ [+0.1cm]
\textbf{Average} & & \textbf{136} & \textbf{273} & \textbf{545} & \textbf{682} & \textbf{818} & \textbf{1363} \\ [+0.1cm]
\hline
        \end{tabular}
        \caption{The translation of our specified $k-$cuts to $\ell-$cut for our different tomographic bins, with the median redshift used to relate them (see Equation~\ref{eqn:Kmax-lmax}). 
        The average over all five tomographic bins is given at the bottom of the table.
        We use the notation $K_{\rm max}$ as this is not a true cut in $k-$space, but rather defined in $k-$space and applied as a cut in $\ell$.}
        \label{tab:ell_max}
\end{table*}

\begin{table*}
    \centering
    
        \begin{tabular}{l|cr|cr|rcc}
        \toprule
            Model Choice &
            \multicolumn{2}{c|}{`Test' Mock}
            & \multicolumn{2}{c|}{`Truth' Mock}
            & \multicolumn{3}{c}{Bias}  \\
        \midrule 
        & $S_8$ & $S_8-S_8^{\rm in}$ &$S_8$ & $S_8-S_8^{\rm in}$ &$ \Delta S_8$ & $\sigma_{\rm Test}/\sigma_{\rm Truth}$ & $\sigma/\sigma_{\rm Fid}$\\ [0.2cm] \midrule 
        Fiducial& $ 0.802^{+ 0.005}_{- 0.006} $ & $ 0.31 \sigma $ & $ 0.799^{+ 0.005}_{- 0.006} $ & $ -0.14 \sigma $ & $ 0.45 \sigma $ & $ 1.04 $ & $ 1.00 $ \\ [+0.1cm] 
        \hline 
        Varying IA model: TATT& $ 0.801^{+ 0.006}_{- 0.006} $ & $ 0.25 \sigma $ & $ 0.799^{+ 0.005}_{- 0.006} $ & $ -0.20 \sigma $ & $ 0.45 \sigma $ & $ 1.04 $ & $ 1.05 $ \\ [+0.1cm] 
        \hline 
        Sensitivity to informative prior width: $h_0$ \& $n_s$& $ 0.802^{+ 0.006}_{- 0.006} $ & $ 0.38 \sigma $ & $ 0.799^{+ 0.005}_{- 0.006} $ & $ -0.12 \sigma $ & $ 0.50 \sigma $ & $ 1.03 $ & $ 1.03 $ \\ [+0.1cm] 
        Varying sum of neutrino mass& $ 0.799^{+ 0.005}_{- 0.005} $ & $ -0.28 \sigma $ & $ 0.796^{+ 0.005}_{- 0.006} $ & $ -0.76 \sigma $ & $ 0.48 \sigma $ & $ 0.98 $ & $ 0.95 $ \\ [+0.1cm] 
        \hline 

        \hline
        \end{tabular}
        
        \caption{Y1 constraints on $S_8$ changing the intrinsic alignment model and varying the Y1 set of cosmological parameters and priors. The second row shows the impact on $S_8$ from using the TATT model for intrinsic alignments compared to the fiducial case using NLA-$z$, the third row shows the impact of using wider priors for $h_0$ and $n_s$ and finally the last row shows the impact of marginalising over neutrino mass.}
\end{table*}

\begin{table*}
    \centering   
    
        \begin{tabular}{ll|cr|cr|rcc}
        \toprule
            Baryon Input & Analysis Model &
            \multicolumn{2}{c|}{`Test' Mock}
            & \multicolumn{2}{c|}{`Truth' Mock}
            & \multicolumn{3}{c}{Bias}  \\
        \midrule 
        & &$S_8$ & $S_8-S_8^{\rm in}$ &$S_8$ & $S_8-S_8^{\rm in}$ &$ \Delta S_8$ & $\sigma_{\rm Test}/\sigma_{\rm Truth}$ & $\sigma/\sigma_{\rm Fid}$\\ [0.2cm] \midrule 
Fiducial & HMCode2020& $ 0.802^{+ 0.005}_{- 0.006} $ & $ 0.31 \sigma $ & $ 0.799^{+ 0.005}_{- 0.006} $ & $ -0.14 \sigma $ & $ 0.45 \sigma $ & $ 1.04 $ & $ 1.00 $ \\ [+0.1cm] 
        \hline 
Strong Feedback & HMCode2020& $ 0.795^{+ 0.006}_{- 0.005} $ & $ -0.83 \sigma $ & $ 0.794^{+ 0.006}_{- 0.007} $ & $ -1.03 \sigma $ & $ 0.20 \sigma $ & $ 0.89 $ & $ 0.97 $ \\ [+0.1cm] 
Weak Feedback & HMCode2020& $ 0.803^{+ 0.005}_{- 0.006} $ & $ 0.59 \sigma $ & $ 0.801^{+ 0.005}_{- 0.006} $ & $ 0.23 \sigma $ & $ 0.36 \sigma $ & $ 1.04 $ & $ 0.93 $ \\ [+0.1cm] 
        \hline 
Fiducial & SP(k)& $ 0.801^{+ 0.005}_{- 0.005} $ & $ 0.30 \sigma $ & $ 0.802^{+ 0.005}_{- 0.005} $ & $ 0.40 \sigma $ & $ -0.11 \sigma $ & $ 0.91 $ & $ 0.85 $ \\ [+0.1cm] 
Fiducial - all scales & SP(k)& $ 0.801^{+ 0.005}_{- 0.005} $ & $ 0.27 \sigma $ & $ 0.798^{+ 0.004}_{- 0.004} $ & $ -0.41 \sigma $ & $ 0.68 \sigma $ & $ 1.19 $ & $ 0.84 $ \\ [+0.1cm] 
Strong Feedback & SP(k)& $ 0.794^{+ 0.005}_{- 0.006} $ & $ -1.11 \sigma $ & $ 0.795^{+ 0.005}_{- 0.005} $ & $ -0.86 \sigma $ & $ -0.25 \sigma $ & $ 1.04 $ & $ 0.96 $ \\ [+0.1cm] 
Strong Feedback - all scales & SP(k)& $ 0.795^{+ 0.005}_{- 0.004} $ & $ -1.14 \sigma $ & $ 0.792^{+ 0.005}_{- 0.005} $ & $ -1.86 \sigma $ & $ 0.72 \sigma $ & $ 0.93 $ & $ 0.79 $ \\ [+0.1cm] 
Weak Feedback & SP(k)& $ 0.803^{+ 0.005}_{- 0.005} $ & $ 0.60 \sigma $ & $ 0.804^{+ 0.005}_{- 0.005} $ & $ 0.71 \sigma $ & $ -0.11 \sigma $ & $ 0.97 $ & $ 0.90 $ \\ [+0.1cm] 

        \hline
        \end{tabular}
        
        \caption{Y1 Constraints on $S_8$ using different baryon feedback simulations in the mock and two different methods for modelling, \hmcode and \spk. 
        We additionally compare the constraints when using all angular scales up to $\ell{\rm max}=5000$ with cuts at $K_{\rm max}=0.5$, our fiducial choice. 
        } 
        \label{tab:baryon_detail}
\end{table*}

\end{document}